\documentclass{article}

\usepackage{arxiv}

\usepackage[utf8]{inputenc} 

\usepackage{amsfonts}       
\usepackage{nicefrac}       
\usepackage{microtype}      
\usepackage{lipsum}

\usepackage{color}
\usepackage{listings}
\usepackage{booktabs}
\usepackage{array}
\usepackage{graphicx}
\usepackage{longtable}

\usepackage{cite}
\usepackage{url}
\usepackage{fancyhdr}

\usepackage{soul}

\usepackage{float}
\usepackage{listings}
\usepackage{mdwmath}
\usepackage{mdwtab}
\usepackage{multirow}
\usepackage{multicol}
\usepackage{rotating}
\usepackage{setspace}
\usepackage[utf8]{inputenc}
\usepackage{lineno}
\usepackage{listings}

\usepackage{mdwmath}
\usepackage{mdwtab}
\usepackage{multirow}
\usepackage{multicol}
\usepackage{array}
\usepackage{booktabs}

\usepackage{enumitem}
\usepackage{xspace}
\usepackage[export]{adjustbox}
\usepackage{graphicx}
\usepackage{color,soul}
\usepackage{rotating}
\usepackage{setspace}
\usepackage{amsmath} 
\usepackage{amssymb}
\usepackage{float}
\usepackage{xcolor}

\usepackage{hyperref}
\usepackage[numbers]{natbib}
\usepackage{url}

\title{A Systematic Literature Review on \\Blockchain Governance}

\author{Yue Liu, Qinghua Lu, Liming Zhu, Hye-Young Paik, Mark Staples\\
Data61, CSIRO, Australia\\
University of New South Wales, Australia\\
yue.liu@data61.csiro.au, qinghua.lu@data61.csiro.au, liming.zhu@data61.csiro.au,\\
h.paik@unsw.edu.au, mark.staples@data61.csiro.au}

\begin{document}

\maketitle

\begin{abstract}
Blockchain has been increasingly used as a software component to enable decentralisation in software architecture for a variety of applications. Blockchain governance has received considerable attention to ensure the safe and appropriate use and evolution of blockchain, especially after the Ethereum DAO attack in 2016. However, there are no systematic efforts to analyse existing governance solutions. To understand the state-of-the-art of blockchain governance, we conducted a systematic literature review with 37 primary studies. The extracted data from primary studies are synthesised to answer identified research questions. The study results reveal several major findings: 1) governance can improve the adaptability and upgradability of blockchain, whilst the current studies neglect broader ethical responsibilities as the objectives of blockchain governance; 2) governance is along with the development process of a blockchain platform, while ecosystem-level governance process is missing, and; 3) the responsibilities and capabilities of blockchain stakeholders are briefly discussed, whilst the decision rights, accountability, and incentives of blockchain stakeholders are still under studied. We provide actionable guidelines for academia and practitioners to use throughout the lifecycle of blockchain, and identify future trends to support researchers in this area.

\end{abstract}

Blockchain, governance, systematic literature review, SLR, distributed ledger technology, DLT

\section{Introduction}

Blockchain, the technology popularised by Bitcoin \citep{Satoshi:bitcoin}, provides decentralised computing and storage infrastructure to build new kinds of trustworthy decentralised applications. A substantial number of projects have been conducted to explore how to use blockchain to increase trust in decentralised settings, without central authorities~\citep{2019-Bratanova-ACS}.

Although blockchain is considered as a viable solution for building decentralised applications, two severe events on Ethereum and Bitcoin have brought concerns to the community that whether the decisions for blockchain development and operation are made in a trustworthy way. In particular, on 18 June 2016, the attack on the Ethereum blockchain's Decentralised Autonomous Organisation (DAO), a decentralised application, caused a loss of over 60 million dollars due to a flaw in smart contract code. This was remedied via a hard fork to reverse the effect of transactions in the attack \citep{DAOattack}. Another event is the dispute on the block size of Bitcoin, which has lasted from 15 August 2015 to 23 January 2016 and resulted in the split of Bitcoin community and ecosystem via forking \citep{selected20}.

These crises of blockchain have provoked the study of blockchain governance and led to a rapid growth of publications. However, there is a lack of systematic literature review to understand the state-of-the-art of blockchain governance. Also, the governance of blockchain often involves a large number of participants due to the decentralised nature of blockchain platforms, which calls for software engineering knowledge to operationalise the governance mechanisms. Therefore, we perform a systematic literature review on existing blockchain governance studies to provide an up-to-date and holistic view of this research topic. We focus on ``blockchain governance", where the object being governed is blockchain itself. Instead of ``governance through blockchain", in which blockchain acts as a means to facilitate governance of software systems.

The whole study followed Kitchenham's guidelines \citep{keele2007guidelines}. The main contributions of this paper are as follows:

\begin{itemize}

    \item We identify a set of {37} primary studies related to blockchain governance published from 2008 to 3 November 2020. Community can use this set of studies as a starting point to conduct further research on blockchain governance. 
    
    \item We present a comprehensive qualitative and quantitative synthesis via identified research questions, reflecting the state-of-the-art in blockchain governance with data extracted from the selected studies. Our synthesis covers the following aspects: definition, motivations, objects, process, stakeholders, and mechanisms of blockchain governance.

    \item We provide insights and research agenda based on our project experiences and SLR results to support further research in the area, which includes both high-level governance principles and actionable governance practices.
\end{itemize}

The remainder of this paper is organised as follows. Section \ref{relatedwork} introduces the background knowledge of blockchain and discusses related work. Section \ref{methodology} introduces the methodology to conduct this study. Section \ref{results} presents the results and our insights. Section \ref{agenda} provides research agenda for the future studies of this topic. Section \ref{threats} analyses the threats to validity of this study. Section \ref{conclusion} concludes the paper.

\section{Background and Related Work}
\label{relatedwork}

\subsection{Blockchain and Smart Contract}

Essentially, blockchain can be considered as a type of distributed ledger technology that can verify and store digital transactions with no central authorities to enhance trust between the interoperating parties \citep{scheuermann2015iacr}. In a blockchain ecosystem, actions within the blockchain platform itself are called ``on-chain", while the others are regarded as ``off-chain" activities. All participants need to make agreement on the on-chain data states during transaction inclusion and confirmation to achieve trust. Nakamoto assumed that the majority of blockchain nodes are honest to reach consensus through game-theoretic incentives, without a third-party intermediary \citep{Satoshi:bitcoin}. 

The varying data states in a blockchain are carried by identifiable transactions which are contained in the blocks. A list of blocks is linked chronologically to form a chain as the ledger of transactional data. In addition, the perception of ``smart contract" improved the computation capability of blockchain networks. Participants can develop customised script, then deploy and execute the smart contract on-chain for differential purposes \citep{Omohundro:2014}. After deployment, a smart contract can express triggers, conditions to enable complex business logic.

Currently, blockchain networks can be categorised into two main groups according to participation permission: permissionless and permissioned blockchains. Permissionless blockchains usually aim to provide a free trading market and thus accept unconditional participation. They have a closed relationship to decentralised finance, where blocks are appended by the elected validators (aka. miners) along with the issuance of on-chain tokens as cryptocurrencies. Permissioned blockchains require acceptance from certain authorities on the participation into the network. This kind of blockchain systems is usually hosted by one or several cooperating organisations.

As an emerging technology, blockchain has gain the attraction of researchers in recent years to exploit the features of blockchain itself. For instance, the scalability of blockchain still remains as a crucial problem that many blockchain platforms (especially permissionless ones) are finding solution tackling low-efficiency issues as low throughput and high transaction latency~\citep{blockchain_scalability}. Meanwhile, many other studies are exploring application domains that use blockchain as a viable infrastructure to set up a decentralised environment. Particularly, there are general works reviewing smart contracts~\citep{smart_contract_slr} and different blockchain applications~\citep{blockchain_application_slr}, and also studies diving into specific usage domains such as IoT~\citep{iot_slr_1, iot_slr_2}, supply chain~\citep{supply_chain_slr_1, supply_chain_slr_2}, healthcare~\citep{healthcare_slr}, agriculture~\citep{agriculture_slr}, smart city~\citep{smart_city_slr}, etc.

\subsection{Related Work}

When selecting the primary studies, we found several papers that conducted survey or review related to blockchain governance. The rest of this section summarises the overview and differences of these papers and our study. Risus and Spohrer \citep{blockchainresearchframework} conducted an SLR to form a research framework. The management and organisation of blockchain were analysed in this framework in terms of users and society, intermediaries, platforms, firms and industries. Politou et al. \citep{blockchainmutability} focused on the mutability of blockchain, claiming the immutable on-chain records may invade individuals' privacy according to related laws and regulations (e.g. GDPR’s ``Right to be Forgotten"). They divided the trade-off approaches of blockchain immutability and regulations into two main groups of redactable blockchain and storage. Kher et al. \citep{ICO} introduced the cryptocurrencies and the events of initial coin offerings, while Zachariadis et al. \citep{ZACHARIADIS2019105} also reviewed the governance issues of blockchain in financial services. Ziolkowski et al. \citep{oldwine} analysed both governance of blockchain and governance through blockchain in terms of six decision problems (i.e., demand and data management, architecture design and development, membership, ownership disputes, and transaction reversal), and the emphasis of governance was placed on four application domains of cryptocurrency, intellectual property rights management, land registries, and supply chain. Smit et al. \citep{decisionrightsgovernance} discussed the decision rights and decision-making process in the governance of blockchain. Li et al. \citep{chinablockchaingovernance} presented the international specification of blockchain, particularly focusing on the standardisation of China's progress.

Compared to the existing studies, our contribution is different in terms of three main aspects (i.e., time frames, methodology, and scope). First, the related work covered relevant studies in this domain from both industry and academia, however, most of them did not provide the state-of-art information. The search time frame in this SLR is 2008-2020 and the selection set results in 2016-2020 (excluding the papers published in Nov to Dec 2020). Secondly, this study adheres to Kitchenham's standard guideline \citep{keele2007guidelines} as the other two SLR, while some reviews had no clear or customised methodology, which may cause bias in data extraction and analysis. Finally, this SLR not only discussed blockchain governance in terms of four objects of blockchain ecosystem (i.e., data, platform, application, community), while also provided the result of data synthesis from six aspects (i.e., what, why, where, when, who, how), which presented a comprehensive and multidimensional perception on the governance of blockchain.

\begin{figure*}[t]
	\centering
	\includegraphics[width=\textwidth]{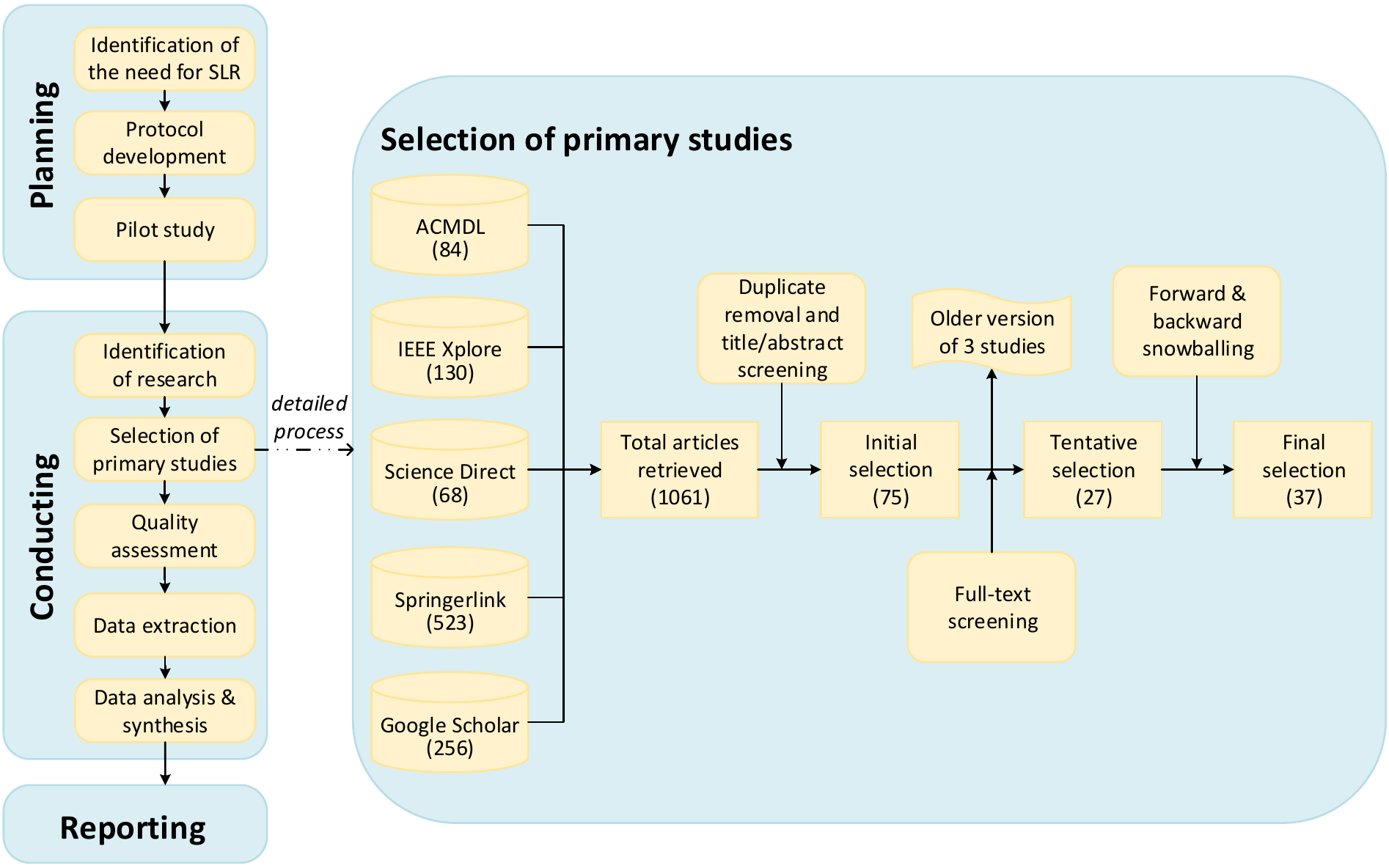}
	\caption{Systematic literature review steps.}
	\label{protocol}
\end{figure*}

\section{Methodology}
\label{methodology}
In this section, we introduce the methodology of this systematic literature review, which follows Kitchenham's guidelines~\citep{keele2007guidelines}. Fig. \ref{protocol} illustrates the steps of this study.

\subsection{Identification of the Need for SLR}

This section describes the motivations of this study.

\subsubsection{Crises in Blockchain}

The topic of blockchain governance has attracted the interests of researchers after the negative events mentioned above. In particular, the DAO attack in Ethereum seriously damaged stakeholders' interests as real money was stolen due to software bugs~\citep{DAOattack}. {On one hand, the attack raised concerns to blockchain's security that whether it is trustworthy to preserve users' data and assets. On the other hand, conducting a hard fork to reverse impacted transactions caused the discussion of whether to preserve the immutability of blockchain under the circumstance that people's money is stolen. Similarly, debate on Bitcoin's blocksize is centered on whether to insist Nakamoto's original design}~\citep{selected20}. {The result of two separate Bitcoin forks has influenced people's confidence that the blockchain community may be fragile. Research interests on governance of blockchain has been boosted by the discussion on these two events.}

\subsubsection{IT Governance and Blockchain Governance}

{Another motivation of this study is that conventional IT governance may not be applicable in the blockchain context. According to Weill}~\citep{ITgovernance}, {IT governance \textit{``specifies the decision rights and accountability framework to encourage desirable behavior in the use of IT".} IT governance is usually applied to better utilise information technology and make profits by providing high-level governance principles and structures within a single organisation. IT governance includes several archetypal approaches, ranging from centralisation to decentralisation. In \textit{business monarchy} and \textit{IT monarchy}, IT-related decisions are made by business executives and IT executives respectively. \textit{Federal system} and \textit{IT duopoly} introduce other executives and business representatives into the decision-making process. A \textit{feudal system} allows each business unit to make separate decisions, whilst individuals can pursue their own IT objectives in \textit{anarchy}. In addition, Weill and Ross} \citep{weill2005matrixed} {propose three IT governance methods: \textit{decision-making structures} help form decision-making committees within an organisation, \textit{alignment processes} clarify the management techniques in governance decisions and implementation (e.g. investment proposal, exception handling), and \textit{formal communications} provide several ways for involving entities to understand the decisions.} 

{In general, IT governance provides high-level guidance for single organisation to better leverage information technology and make profits. Nevertheless, there is still a lack of holistic view to facilitate operational blockchain governance. Blockchain is hard to govern as the code is stand-alone and difficult to update. The decentralised nature of blockchain requires the participation of majority or even all stakeholders for an update decision to achieve democracy. Consequently, we decide to conduct an SLR to learn the state-of-art of blockchain governance, from which we can extract the actionable mechanisms from current progress in academia.}

\begin{table*}[h!]
\footnotesize
\centering
\caption{Research questions and motivations.}
\label{question}
\begin{tabular}{p{0.04\columnwidth}p{0.35\columnwidth}p{0.46\columnwidth}}
\toprule

\textbf{No} & \textbf{Research question} & \textbf{Motivation}\\
\midrule

\multirow{2}{0.04\columnwidth}{RQ1} & \multirow{2}{0.35\columnwidth}{What is blockchain governance?} & {To observe how primary studies define the concept of blockchain governance.}\\
\cmidrule(l){1-3}

\multirow{2}{0.04\columnwidth}{RQ2} & \multirow{2}{0.35\columnwidth}{Why is blockchain governance adopted?} &  To understand the challenges that blockchain governance aims to address.\\
\cmidrule(l){1-3}

\multirow{2}{0.04\columnwidth}{RQ3} & \multirow{2}{0.35\columnwidth}{Where is blockchain governance enforced?} & {To distinguish the key governance objects in blockchain ecosystem.}\\
\cmidrule(l){1-3}

\multirow{1}{0.04\columnwidth}{RQ4} & \multirow{1}{0.35\columnwidth}{When is blockchain governance applied?} & {To understand the whole process of blockchain governance.}\\
\cmidrule(l){1-3}

\multirow{2}{0.04\columnwidth}{RQ5} & \multirow{2}{0.35\columnwidth}{Who is involved in blockchain governance?} & To identify the different roles, and their authorities, capabilities and responsibilities in blockchain governance.\\
\cmidrule(l){1-3}

\multirow{2}{0.04\columnwidth}{RQ6} & \multirow{2}{0.35\columnwidth}{How is blockchain governance designed?} & {To explore actionable mechanisms for implementing blockchian governance.}\\

\bottomrule
\end{tabular}
\end{table*}

\subsection{Protocol Development and Pilot Study}
{In this section, we mainly introduce the research questions in our protocol, while the inclusion and exclusion criteria, and quality assessment in subsequent sections.} We adopt 5W1H (what, why, where, when, who, and how) as the research questions in our SLR protocol, to holistically examine the adoption, use and design of blockchain governance. Table \ref{question} lists the research questions addressed by this study and their motivations. 
{To validate the developed protocol, we conduct a pilot study in which the six research questions are applied to five papers to extract answers. The papers are selected by searching ``blockchain governance" via Google Scholar, including}~\citep{selected11, selected14, selected7, selected5, selected29}. {The pilot study results indicate that our protocol is feasible.}

\begin{table*}[tbhp]
\footnotesize
\centering
\caption{Paper selection results.}
\label{selection}
\begin{tabular}{p{0.1\columnwidth}p{0.1\columnwidth}p{0.1\columnwidth}p{0.1\columnwidth}p{0.1\columnwidth}p{0.1\columnwidth}p{0.1\columnwidth}}
\toprule

\multicolumn{1}{l}{\multirow{2}{0.1\columnwidth}{\bf Sources}} &
\multicolumn{1}{c}{\multirow{2}{0.1\columnwidth}{\bf ACM}} &
\multicolumn{1}{c}{\multirow{2}{0.1\columnwidth}{\bf IEEE Xplore}} &
\multicolumn{1}{c}{\multirow{2}{0.1\columnwidth}{\bf Science Direct}} &
\multicolumn{1}{c}{\multirow{2}{0.1\columnwidth}{\bf Springer Link}} &
\multicolumn{1}{c}{\multirow{2}{0.1\columnwidth}{\bf Google Scholar}} &
\multicolumn{1}{c}{\multirow{2}{0.1\columnwidth}{\bf Total}}\\ \\
\midrule

\bf{Count} & {3} & {4} & {1} & {{8}} & {{21}} & {{37}}\\

\bottomrule
\end{tabular}
\end{table*}

\begin{table}[tbhp]
\footnotesize
\centering
\caption{Search terms.}
\label{terms}
\begin{tabular}{p{0.2\columnwidth}p{0.5\columnwidth}}
\toprule

\multicolumn{1}{l}{\bf Key Term} &
\multicolumn{1}{c}{\bf Supplementary Terms}\\
\midrule

Blockchain & {Distributed ledger technology, DLT}\\
\cmidrule(l){1-2}

Governance & {Govern, Governing}\\

\bottomrule
\end{tabular}
\end{table}

\begin{table*}[hptb]
\footnotesize
\centering
\caption{Search strings and quantity of \textit{ACM Digital Library}.}
\label{ACM}
\begin{tabular}{p{0.2\columnwidth}p{0.7\columnwidth}}
\toprule

\multirow{5}{0.2\columnwidth}{\bf Search string} & \texttt{(Title:(blockchain OR "distributed ledger technology" OR DLT) AND Title:(governance OR governing OR govern)) OR (Abstract:(blockchain OR "distributed ledger technology" OR DLT) AND (Abstract:(governance OR governing OR govern))} \\
\cmidrule(l){1-2}

\multirow{1}{0.2\columnwidth}{\bf Result quantity} & {84}\\
\cmidrule(l){1-2}

\multirow{1}{0.2\columnwidth}{\bf Selected papers} & {\bf 3}\\

\bottomrule
\end{tabular}
\end{table*}

\begin{table*}[hptb]
\footnotesize
\centering
\caption{Search strings and quantity of {\textit{IEEEXplore}}.}
\label{IEEE}
\begin{tabular}{p{0.2\columnwidth}p{0.7\columnwidth}}
\toprule

\multirow{5}{0.2\columnwidth}{\bf Search string} & \texttt{(("Document Title":blockchain OR "distributed ledger technology" OR DLT) AND ("Document Title":governance OR governing OR govern)) OR (("Abstract":blockchain OR "distributed ledger technology" OR DLT) AND ("Abstract":governance OR governing OR govern))} \\
\cmidrule(l){1-2}

\multirow{1}{0.2\columnwidth}{\bf Result quantity} & {130}\\
\cmidrule(l){1-2}

\multirow{1}{0.2\columnwidth}{\bf Selected papers} & {\bf 4}\\

\bottomrule
\end{tabular}
\end{table*}

\begin{table*}[hptb]
\footnotesize
\centering
\caption{Search strings and quantity of \textit{ScienceDirect}.}
\label{ScienceDirect}
\begin{tabular}{p{0.2\columnwidth}p{0.7\columnwidth}}
\toprule

\multirow{3}{0.2\columnwidth}{\bf Search string} & \texttt{Title, abstract, keywords: (blockchain OR "distributed ledger technology" OR DLT) AND (governance OR governing OR govern)} \\
\cmidrule(l){1-2}

\multirow{1}{0.2\columnwidth}{\bf Result quantity} & {68}\\
\cmidrule(l){1-2}

\multirow{1}{0.2\columnwidth}{\bf Selected papers} & {\bf 1}\\

\bottomrule
\end{tabular}
\end{table*}

\begin{table*}[hptb]
\footnotesize
\centering
\caption{Search strings and quantity of \textit{SpringerLink}.}
\label{Springer}
\begin{tabular}{p{0.2\columnwidth}p{0.7\columnwidth}}
\toprule

\multirow{2}{0.2\columnwidth}{\bf Search string} & \texttt{(blockchain OR "distributed ledger technology" OR DLT) NEAR/10 (governance OR governing OR govern)} \\
\cmidrule(l){1-2}

\multirow{1}{0.2\columnwidth}{\bf Result quantity} & {523}\\
\cmidrule(l){1-2}

\multirow{4}{0.2\columnwidth}{\bf Remark} & The search engine of \textit{SpringerLink} does not support composite logical operators for title, hence, we decided to search papers in which the word "governance" is closed to "blockchain" within 10 words.\\
\cmidrule(l){1-2}

\multirow{1}{0.2\columnwidth}{\bf Selected papers} & {\bf {8}}\\

\bottomrule
\end{tabular}
\end{table*}

\begin{table*}[hptb]
\footnotesize
\centering
\caption{Search strings and quantity of \textit{Google Scholar}.}
\label{Google}
\begin{tabular}{p{0.2\columnwidth}p{0.7\columnwidth}}
\toprule

\multirow{2}{0.2\columnwidth}{\bf Search string} & \texttt{allintitle: (blockchain OR "distributed ledger technology" OR DLT) (governance OR governing OR govern)} \\
\cmidrule(l){1-2}

\multirow{1}{0.2\columnwidth}{\bf Result quantity} & {256}\\
\cmidrule(l){1-2}

\multirow{2}{0.2\columnwidth}{\bf Remark} & Search title only (\textit{Google Scholar} does not support abstract or keyword search option).\\
\cmidrule(l){1-2}

\multirow{1}{0.2\columnwidth}{\bf Selected papers} & {\bf {21}}\\

\bottomrule
\end{tabular}
\end{table*}

\subsection{Identification of Research}

``Blockchain governance" was used as the key word, and we also added synonyms, abbreviations as supplements to include more comprehensive search results, which are shown in Table \ref{terms}. For ``blockchain", many studies relate cryptocurrencies to this technology and may replace each other in certain cases. In this research, we focus on the blockchain technology itself, and treat cryptocurrency as an application based on it, thus the supplementary terms do not contain any crypto-related words. For ``governance", we only select two cognates to avoid the ambiguity and misunderstanding in other synonyms.

We collected the literature from the following sources: 1) \textit{ACM Digital Library}, 2) {\textit{IEEEXplore}}, 3) \textit{ScienceDirect}, 4) \textit{SpringerLink}, and 5) \textit{Google Scholar}. {The selection of data sources was driven by the willingness of searching as many papers as possible to properly perform this systematic literature review. In this regard, the above selected data sources are all recognised to be the most representative digital libraries for Software Engineering research}~\citep{keele2007guidelines}. The search time frame is set from 1 January 2008 (as the concept of blockchain is {popularised} in the Whitepaper of Bitcoin \citep{Satoshi:bitcoin}) to 3 November 2020. We designated the search strings for each source for paper collection. Each search string was tested for its appropriateness and effectiveness, and the final search terms for the five source are presented in Table \ref{ACM} to \ref{Google}.

The original search resulted in 1061 papers: 84 from \textit{ACM Digital Library}, 130 from {\textit{IEEEXplore}}, 68 from \textit{ScienceDirect}, 523 from \textit{SpringerLink}, and 256 from \textit{Google Scholar}. After the removal of duplicates, and first inclusion/exclusion from title and abstract, the initial selection set had 75 papers. Then, we conducted a full text screening and ended up with 27 papers as the tentative selection. After the following snowballing phase, {10} papers were included and the final selection set has {37} papers. The number of papers per source are presented in Table \ref{selection}.

\subsection{Selection of Primary Studies}
\label{criteria}
We formulated the inclusion and exclusion criteria to conduct selection on collected papers. The drafted criteria were first improved with the pilot study of five studies. Afterwards, four researchers reviewed and updated the criteria. Seven criteria were finalised to determine the eligibility of selected papers. The inclusion criteria are as follows.

\begin{itemize}
    \item {A paper that proposes a solution for governance of blockchain.}
    
    \item {A paper that proposes principles or frameworks for developing governance of blockchain.}
\end{itemize}

The exclusion criteria are as follows.

\begin{itemize}
    \item Papers that focus on ``governance through blockchain" instead of ``governance of blockchain". 
    
    \item {Older version of a study that has a more comprehensive version.}
    
    \item Papers that are not written in English.
    
    \item {Papers that are not accessible.}
    
    \item {Survey, review and SLR papers. These studies are selected and identified in a separate subset, to provide better understanding of state-of-art on blockchain governance. However, we do not conduct data extraction or synthesis from these studies as they are considered as the related work of this study.}
\end{itemize}

Afterwards, we carried out a forward and backward snowballing process, to include any other related studies that might have been missed in the initial search. {Hereby, backward and forward snowballing refers to using the references and citations of a paper respectively to find related studies}~\citep{snowballing}. {Ten} {studies were included after the snowballing process, which are illustrated in Table} \ref{tab:snowballing}.

The overall selection process involved two researchers. A researcher first included and excluded the papers from title and abstract, the results were then being screened for full text. The other researcher was responsible for the review of the results in both phases. Two other researchers were consulted when disagreements happened. The whole collection and selection process is demonstrated in Fig.~\ref{protocol}.

\begin{table*}[hptb]
\footnotesize
\centering
\caption{{Snowballing result.}}
\label{tab:snowballing}
\begin{tabular}{p{0.2\columnwidth}p{0.25\columnwidth}}
\toprule

{\bf Seed paper} & {\bf Snowballed paper} \\
\cmidrule(l){1-2}

\citep{selected4} & \citep{selected20} \\
\cmidrule(l){1-2}

\citep{selected7} & \citep{selected30, selected34} \\
\cmidrule(l){1-2}

\citep{selected19} & \citep{selected3} \\
\cmidrule(l){1-2}

\citep{selected20} & \citep{selected12, selected26, selected33, selected36, selected37} \\
\cmidrule(l){1-2}

\citep{selected34} & \citep{selected16} \\

\bottomrule
\end{tabular}
\end{table*}

\subsection{Quality Assessment}
\leavevmode\\
\indent We conducted an assessment to the selected papers to {evaluate their quality and finalise} the inclusion eligibility~\citep{keele2007guidelines, quality_assessment}. {Five quality criteria (QC) were developed and each one can be answered by one of two scores: $1$ (yes), $0.5$ (partially), and $0$ (no). The total scores of five QC were calculated and ranked the papers into three categories: good (4 $\textless$ \textit{score} $\leq$ 5), fair (2 $\textless$ \textit{score} $\leq$ 4), and fail (0 $\leq$ \textit{score} $\leq$ 2). The studies in ``good" and ``fair" groups were included while the ``fail" papers are excluded in this phase. During the conduction of quality assessment, the {37} papers were all selected as the results are all higher than three scores, which indicates that the selected studies all have fair or good quality. The QC are as follows:}

\begin{itemize}
    \item QC1. Does the study identify a main research purpose? A primary study should have a clear purpose on the investigation of blockchain governance.
    \item QC2. Does the study clearly define the concept of blockchain governance? The definition can help develop a comprehensive perception of blockchain governance.
    \item QC3. Does the study describe the methodology clearly? The applied methodology can embody the relevance of a paper to this research.
    \item QC4. Does the study conduct an evaluation? Are the proposed solutions solid and how to evaluate them? These factors are helpful to design available and applicable governance methods.
    \item QC5. Does the study discuss any limitations? The discussion on current limitations of blockchain governance can reveal the direction of future studies on this research topic.
\end{itemize}

\subsection{Data Extraction and Synthesis}
\leavevmode\\
\indent We downloaded all selected papers from the search sources for data extraction. A Google sheet was developed to record original data from the papers, including basic information (i.e., title, publication year, search source, type, venue, authors, and affiliation), and the answers of each RQ, while the calculation of QC scores is also included in this sheet. The basic information of primary studies are presented in Appendix.

Four researchers participated in this phase. First, two researchers are responsible to extract data, and conduct cross-validation to each other's extraction. If there were disagreements on certain data, another two independent researchers were consulted to review the paper and finalise the extraction decisions.


\section{Results}
\label{results}


In this section, we summarise the extracted results of {each research question}.

\subsection{RQ1: What is blockchain governance?}
\label{section-what}

\begin{figure*}[t]
	\centering
	\includegraphics[width=0.85\textwidth]{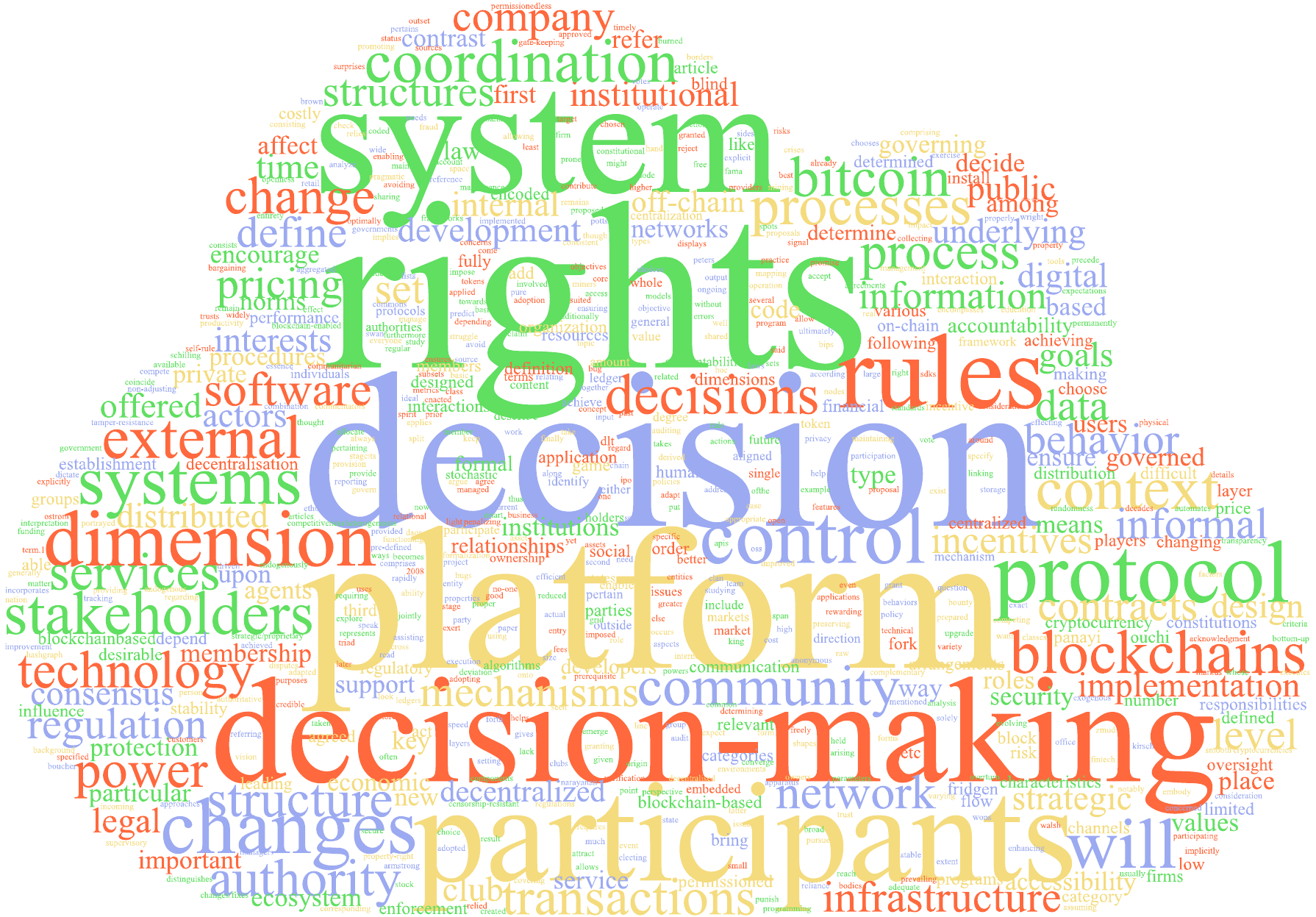}
	\caption{RQ1: What is blockchain governance?}
	\label{what}
\end{figure*}

{The first research question is ``What is blockchain governance?" The motivation is to observe how literature define the concept of blockchain governance. To answer this question, the definition of blockchain governance given by each study is recorded. Fig.} \ref{what} {is a word cloud showing the frequency of the words, which appear in the original definition of blockchain governance in each study. The top ten appeared words include: \textit{right} (27), \textit{decision} {(25)}, \textit{rules} {(24)}, \textit{platform} (23),  \textit{system} (19), \textit{control} (18), \textit{decision-making} {(17)}, \textit{participants} {(16)}, \textit{bitcoin} (13), \textit{community} (13).} {Further, we found that seven studies adopt theories of governance from other domains, including IT governance} \citep{selected11,selected12,selected14}, {corporate governance} \citep{selected7, selected19}, {complex adaptive systems} \citep{selected10}, {and formal political economy} \citep{selected24}.

\begin{table*}[!h]
\footnotesize
\centering
\caption{{Dimensions of blockchain governance definition.}}
\label{dimension}
\begin{tabular}{p{0.19\columnwidth}p{0.2\columnwidth}p{0.51\columnwidth}}
\toprule

{\bf Categories} & {\bf Dimensions} & {\bf Primary studies} \\
\cmidrule(l){1-3}

\multicolumn{1}{c}{\multirow{5}{0.19\columnwidth}{\bf IT governance}} & \multirow{2}{0.23\columnwidth}{Decision rights} & \citep{selected5, selected7, selected8, selected10, selected11, selected12, selected13, selected14, selected15, selected17, selected19, selected22, selected23, selected25, selected26, selected30, selected31, selected32, selected34, selected35}\\
\cmidrule(l){2-3}

& Accountability & \citep{selected5, selected13, selected14} \\
\cmidrule(l){2-3}

& Incentives & \citep{selected2, selected4, selected11, selected12, selected13, selected14, selected19, selected20, selected37}\\
\cmidrule(l){1-3}

\multicolumn{1}{c}{\multirow{2.5}{0.19\columnwidth}{\bf Decentralisation level}} & Permissioned & \citep{selected1, selected31} \\
\cmidrule(l){2-3}

& Permissionless & \citep{selected31} \\
\cmidrule(l){1-3}

\multicolumn{1}{c}{\multirow{2.5}{0.19\columnwidth}{\bf Governance context}} & On-chain & \citep{selected21, selected22, selected27, selected32, selected37} \\
\cmidrule(l){2-3}

& Off-chain & \citep{selected21, selected22, selected27, selected32, selected37} \\

\bottomrule
\end{tabular}
\end{table*}

\begin{figure*}[h!]
\begin{center}
\setlength{\fboxrule}{0.8pt}
\noindent

\fbox{%
\begin{minipage}{\textwidth}
\vspace{0.7em}
\footnotesize
\begin{center}
    \textbf{Insight from RQ1: What is blockchain governance?}\\
\end{center} 

\begin{quotation}

{\textit{\textbf{Definition of blockchain governance:}} Since there is not a commonly recognised definition of blockchain governance, hereby, we give a high-level definition based on the extracted information:}

\textit{\textbf{{``Blockchain governance refers to the structures and processes that are designed to ensure the development and use of blockchain are compliant to legal regulations and ethical responsibilities."}}}

{Hereby, ``structures" represent the architecture design of a blockchain, while ``processes" denote the overall development process.}

\vspace{0.5em}

{\textit{\textbf{Dimensions of blockchain governance:}} Blockchain consists of the three dimensions of decision rights, accountability, and incentives, whereby they should all align with the decentralisation level of deployed blockchain.} First, decision rights of an entity may vary throughout the lifecycle of blockchain. For example, a user can compete to be a block validator, who is capable to select the transactions for inclusion. Secondly, pseudonymous address is a challenge for the identification of accountable entities, particularly in permissionless blockchain. Thirdly, incentives can be realised through on-chain token distribution and off-chain business agreements. 

\end{quotation}
\vspace{0.7em}
\end{minipage}}
\end{center} 
\end{figure*}

{To answer RQ1 more accurately, we summarise three categories from the extracted data and their involved dimensions, as shown in Table}~\ref{dimension}. {First, based on the observation that several studies comprehend blockchain governance from the perspective of IT governance, we recap the three dimensions of IT governance (i.e., decision rights, accountability, and incentives)}~\citep{ITgovernance} {in blockchain governance as follows.}

\begin{itemize}
    \item \textit{Decision Rights}: {Decision rights denote the authority, responsibility, and capability of involved individuals in a blockchain that how decisions are made and monitored}~\citep{selected7, selected11, selected30}, {and which stakeholders' interests should be prioritised}~\citep{selected26}. {The allocation of decision rights can also indicate the degree of decentralisation of a blockchain}~\citep{selected14}. {Ellul et al.} \citep{selected25} {state that no single party can determine the appending transactions, while two other studies also mention all decisions in a blockchain are made by the collective power}~\citep{selected5, selected34}.

     \item \textit{Accountability}: Accountability means that those responsible for the different phases of the blockchain lifecycle should be identifiable and answerable for their decisions and the outcomes of blockchain. {Consequently, this governance dimension can ensure the efficient use of resources and monitor the overall performance of blockchain platform}~\citep{selected5}. {Usually, accountability in blockchain enacted via on-chain smart contracts and off-chain legal agreements}~\citep{selected14}.

    \item \textit{Incentives}: Incentives are considered to be a motivational factor that influences participants’ behaviors~\citep{selected11}, which can be either positive or negative (i.e., rewards for contributions, or sanctions for malicious operations). {On one hand, incentives attract individuals to participate in the governance issues}~\citep{selected12, selected20}. {On the other hand, incentives can guide different stakeholders to make decisions as a whole}~\citep{selected2, selected14, selected37}. 

\end{itemize}

{Secondly, considering the different blockchain types, Werner et al.}~\citep{selected31} {state that the trust and perceived risks are dependent on the decentralisation level and permission to participate a blockchain. Thirdly, {five} studies directly give the definition of on-chain and off-chain governance~}\citep{selected21, selected22, selected27, selected32, selected37}: {1) on-chain governance focuses on the decision-making process that are codified to a blockchain, while subsequent interactions should adhere to these rules of code, and; 2) off-chain governance comprises all other off-chain process that may influence the development and operation of blockchain.}















\begin{table*}[!h]
\footnotesize
\centering
\caption{{RQ2: Why is blockchain governance adopted?}}
\label{why}
\begin{tabular}{p{0.2\columnwidth}p{0.7\columnwidth}}
\toprule

{\bf Categories} & {\bf Primary studies} \\
\cmidrule(l){1-2}

{\bf Adaptability} & \citep{selected5, selected6, selected7, selected8, selected9, selected10, selected13, selected16, selected18, selected20, selected21, selected26, selected28, selected30, selected31, selected33, selected34, selected35, selected37} \\
\cmidrule(l){1-2}

{\bf Upgradability} & \citep{selected2, selected3, selected8, selected9, selected10, selected11, selected20, selected21, selected23, selected25, selected27, selected28, selected32, selected33, selected34, selected36} \\
\cmidrule(l){1-2}

{\bf Security} & \citep{selected4, selected11, selected13, selected16, selected17, selected18, selected22, selected25, selected26, selected28, selected29, selected33, selected34, selected35, selected36, selected37}  \\
\cmidrule(l){1-2}

{\bf Fairness} & \citep{selected1, selected20, selected34, selected37}  \\
\cmidrule(l){1-2}

{\bf Accountability} & \citep{selected15, selected25} \\
\cmidrule(l){1-2}

{\bf Privacy} & \citep{selected26} \\
\cmidrule(l){1-2}

{\bf Censorship-resistance} & \multirow{1}{0.7\columnwidth}{\citep{selected3}} \\

\bottomrule
\end{tabular}
\end{table*}


\subsection{RQ2: Why is blockchain governance adopted?}

The motivation of RQ2 is to understand {the main challenges blockchain governance aims to resolve.} As illustrated in Table \ref{why}, we classify the answers into seven categories: including adaptability, upgradability, security, fairness, accountability, privacy, {and censorship-resistance}~\citep{softwarequality, humanvalue}.  


\begin{itemize}

    \item \textit{Adaptability}: Adaptability is the main motivation of adopting blockchain governance that an effective governance structure can widen the application scenarios of blockchain~\citep{selected5, selected7, selected13, selected26}. {Proper governance can help minimise the risks of different applications, hence blockchain can adapt to specific needs and restrictions of both public and private sectors}~\citep{selected16, selected18}.
    

    \item \textit{Upgradability}: Upgradability is the second major objective to apply governance in blockchain, which means the capability of being upgraded in functionality by adding or replacing blockchain components. {Blockchain intends to provide immutable data storage, nevertheless, such absolute immutability is not desirable as on-chain governance relying on original code has proved to be insufficient}~\citep{selected32}. {Blockchain platform should be upgraded to fix bug and adopt new practices and behaviors}~\citep{selected2, selected3, selected11, selected27}. {Specifically, the Bitcoin block size debate is viewed as an example that governance is needed for upgrading blockchain}~\citep{selected3, selected8, selected20, selected21, selected25}.

    \item \textit{Security}: {Security can ensure that the daily operation of blockchain is in a trustworthy manner, protecting on-chain data and digital assets against malicious attacks}~\citep{selected4}. {A significant aspect is that the vulnerabilities in code should be mitigated}~\citep{selected13}. {Notably, several studies all mention the Ethereum DAO attack, which was caused by a smart contract bug}~\citep{selected22, selected33, selected34, selected35}. {This attack made the entire Ethereum system to stop and ultimately ended up in a hard fork. Such crisis has affected the reputation of blockchain, and intensified the need of blockchain governance to prevent malicious behaviors and ensure the security of whole system. In recent years, blockchain is often associated with data violations and cyber crimes as there is a lack of proper governance to manage legal compliance}~\citep{selected16, selected25, selected26}.

    
    \item \textit{Fairness}: Effective blockchain governance can maintain the fairness of differential parities in the blockchain ecosystem and community. {The blockchain community consists of multiple types of stakeholders, hence, governance models are required to maintain the fairness and social order of these stakeholders regarding the complex realities and variety of human motivations, behaviors, and decision-makings}~\citep{selected1, selected20, selected34, selected37}. 
    

    
    \item \textit{Accountability}: In RQ1, accountability is one of the three dimensions adopt from IT governance, and this is considered as the consequence brought by governance structures. While in RQ2, there are two primary studies identifying this attribute as the motivation of blockchain governance. {Howell et al.} \citep{selected15} {states that neither the on-chain ledger nor platform itself is claimed to be possessed by any entity, however, the responsibility is assumed to be taken by a particular party for the ongoing operation. Ellul et al.} \citep{selected25} {mentions removing on-chain anonymity requires a regulatory framework in the context of decentralised governance.}
    
    
    
    \item \textit{Privacy}: In a blockchain network, data are generally transparent and visible to all nodes that participate in the network and all devices with access to those nodes. This might pose a risk in situations where  sensitive  data  should  only  be  visible  to  selected  participants. Governance is expected to protect sensitive data stored in blockchain~\citep{selected26}.
    
    
    \item \textit{{Censorship-resistance}}: {Nabilou} \citep{selected3} {states that blockchain is a censorship-resistant technique which enables rules as code. Preserving censorship-resistance relies on the collective decisions of all participants. Adopting governance can facilitate the decentralisation in blockchain and hence, retain the censorship-resistant property.}

 \end{itemize}

\begin{figure*}[h!]
\begin{center}
\setlength{\fboxrule}{0.8pt}
\noindent

\fbox{%
\begin{minipage}{\textwidth}
\vspace{0.7em}
\footnotesize
\begin{center}
    \textbf{Insight from RQ2: Why is blockchain governance adopted?}\\
\end{center} 

\begin{quotation}
\textbf{\textit{Ethical values:}} Adaptability and upgradability are the two main properties motivating the governance of blockchain, while some other identified motivations of blockchain governance are related to ethical values, e.g., security, fairness, privacy, which reflects the awareness of blockchain’s impact on human and society. 

\end{quotation}
\vspace{0.7em}
\end{minipage}}
\end{center} 
\end{figure*}

\begin{table*}[!h]
\footnotesize
\centering
\caption{{RQ3: Where is blockchain governance enforced?}}
\label{where}
\begin{tabular}{p{0.2\columnwidth}p{0.7\columnwidth}}
\toprule

{\bf Ecosystem} & \multirow{1}{0.7\columnwidth}{{\bf Primary studies}} \\
\cmidrule(l){1-2}

\multirow{2}{0.2\columnwidth}{\bf Platform} & \citep{selected2, selected3, selected4, selected6, selected7, selected8, selected9, selected10, selected11, selected12, selected13, selected14, selected15, selected16, selected17, selected19, selected20, selected21, selected22, selected23, selected24, selected25, selected26, selected27, selected28, selected29, selected30, selected31, selected33, selected34, selected35, selected36, selected37} \\
\cmidrule(l){1-2}

\multirow{2}{0.2\columnwidth}{\bf Community} & \citep{selected1, selected3, selected4, selected5, selected6, selected7, selected9, selected10, selected11, selected13, selected15, selected16, selected17, selected19, selected20, selected21, selected22, selected23, selected24, selected26, selected27, selected28, selected30, selected31, selected32, selected33, selected34, selected35, selected36, selected37} \\
\cmidrule(l){1-2}

{\bf Application} & \citep{selected8, selected10, selected12, selected13, selected14, selected16, selected18, selected23, selected25, selected26, selected28, selected30, selected31, selected35}\\
\cmidrule(l){1-2}

{\bf Data} & \citep{selected1, selected6, selected13, selected14, selected18, selected26, selected30, selected32}\\

\bottomrule
\end{tabular}
\end{table*}

\subsection{RQ3: Where is blockchain governance enforced?}
\label{section-where}

{The motivation of RQ3 is to distinguish the key governance objects within blockchian ecosystem. We summarise four different governance objects} as shown in Table \ref{where}, including data, platform, application, and community. 

Majority of primary studies focus on the governance of blockchain platform itself {(89\%).} 
{In particular, the governance of blockchain should determine how the platform is designed}~\citep{selected15}, for instance, infrastructure configuration (e.g. block size and interval)~\citep{selected9, selected23, selected36}, and consensus protocol (e.g. transaction generation and confirmation process)~\citep{selected7, selected10, selected19}. Moreover, governance can facilitate the upgrade of platform to meet the new requirements according to users' feedback and proposals~\citep{selected7, selected21, selected23, selected24, selected33}. 

{81\%} {of the primary studies discuss governance of the blockchain community. Violations of on-chain governance may cause conflicts in real-world context}~\citep{selected22}, {consequently, governance of community places an emphasis on both formal and informal governance processes of different stakeholders and their collaborations via off-chain channels (e.g., Twitter, Reddit and GitHub)}~\citep{selected3, selected7, selected19, selected21, selected33}. Governance over community also includes institutional governance which covers the legitimacy of all aspects in the blockchain ecosystem against injustice~\citep{selected10}. The development and business activities associated with blockchain ecosystem should comply with the regulations issued by the states that the blockchain is expected to deploy~\citep{selected16, selected17, selected26}.

14 out of {37} primary studies express that governance is required to be enforced over the blockchain applications, where blockchain is exploited as a component to enable decentralised architecture. Consequently, the usage of blockchain in the applications should be governed according to industry regulations and specifications~\citep{selected23, selected26}. {Specifically, cryptocurrencies are taken as examples that legal solutions should be employed to against criminal activities such as money laundering and terrorist financing}~\citep{selected16, selected25}.

There are 8 primary studies discussing on-chain data governance. In particular, Atzori~\citep{selected6} {focuses on the data source and quality to ensure the conformance of on-chain data. Once a block is validated and appended to the blockchain, all included data is permanently stored unless human interventions are involved to edit or reverse historical transactions}~\citep{selected13, selected18}. {Stored data should be audited to detect illegal actions}~\citep{selected1}. {Violations to regulations such as data protection or child pornography require governance methods to remove relevant transaction records}~\citep{selected32}.

\begin{figure*}[thbp]
\begin{center}
\setlength{\fboxrule}{0.8pt}
\noindent

\fbox{%
\begin{minipage}{\textwidth}
\vspace{0.7em}
\footnotesize
\begin{center}
    \textbf{Insight from RQ3: Where is blockchain governance enforced?}\\
\end{center} 

\begin{quotation}

The majority of blockchain governance studies focus on the design of \textbf{\textit{platform}} and organisation of \textbf{\textit{community}} activities, while less attention is paid to applications and data.

\textbf{\textit{Application:}} The governance of application mostly discusses cryptocurrency, {whereby the influences of other application domains are missing.}

\textbf{\textit{Data:}} Governance over the \textbf{\textit{data}} can be enabled throughout \textbf{\textit{on-chain data lifecycle}}, e.g. checking submitted data through client applications, classifying data based on its degree of sensitivity, adding supernodes with higher level of data access permission.

\end{quotation}
\vspace{0.7em}
\end{minipage}}
\end{center} 
\end{figure*}






\begin{table*}[!h]
\footnotesize
\centering
\caption{{RQ4: When is blockchain governance applied?}}
\label{when}
\begin{tabular}{p{0.2\columnwidth}p{0.7\columnwidth}}
\toprule

{\bf Categories} & {\bf Primary studies} \\
\cmidrule(l){1-2}

\multirow{2}{0.2\columnwidth}{\bf Planning} & \citep{selected2, selected3, selected4, selected5, selected6, selected7, selected8, selected9, selected10, selected11, selected12, selected13, selected14, selected15, selected16, selected17, selected19, selected20, selected21, selected22, selected23, selected24, selected26, selected30, selected31, selected33, selected35, selected36, selected37}\\
\cmidrule(l){1-2}

\multirow{1}{0.2\columnwidth}{\bf Analysis} & \citep{selected2, selected4, selected7, selected8, selected9, selected11, selected12, selected13, selected14, selected15, selected17, selected20, selected21, selected22, selected23, selected30, selected33, selected36, selected37}\\
\cmidrule(l){1-2}

\multirow{1}{0.2\columnwidth}{\bf Design} & \citep{selected2, selected3, selected7, selected9, selected11, selected12, selected13, selected14, selected15, selected18, selected19, selected20, selected22, selected23, selected29, selected30, selected31, selected32, selected34, selected35, selected36}\\
\cmidrule(l){1-2}

\multirow{1}{0.2\columnwidth}{\bf Implementation} & \citep{selected3, selected4, selected8, selected9, selected11, selected12, selected13, selected14, selected15, selected20, selected22, selected24, selected25, selected33, selected34, selected35, selected36} \\
\cmidrule(l){1-2}

\multirow{1}{0.2\columnwidth}{\bf Testing} & \citep{selected11, selected23, selected27, selected28} \\
\cmidrule(l){1-2}

\multirow{2}{0.2\columnwidth}{\bf Operation} & \citep{selected1, selected2, selected3, selected4, selected6, selected7, selected8, selected9, selected11, selected12, selected13, selected14, selected15, selected16, selected18, selected19, selected20, selected22, selected23, selected25, selected29, selected30, selected31, selected32, selected34, selected35, selected36, selected37} \\
\cmidrule(l){1-2}

\multirow{1}{0.2\columnwidth}{\bf Termination} & \citep{selected3, selected4, selected8, selected15, selected22} \\

\bottomrule
\end{tabular}
\end{table*}

\subsection{RQ4: When is blockchain governance applied?}
\label{section when}

The motivation of RQ4 is to understand {the overall process of blockchain governance}. Currently, there is not a commonly recognised lifecycle for the development process of blockchain platforms. {To answer this question, we adopt the software development process}~\citep{softwaredevelopmentprocess} {to analyse and synthesise extracted data, as shown in Table} \ref{when}.

{During the development of a blockchain, governance is instrumentally embedded with the settings and design of on-chain rules, which are considered as ``endogenous governance"}~\citep{selected7}. {In the \textit{planning} phase, the founding members should make institutional arrangements at the first place, for instance, the initial distribution of decision rights}~\citep{selected15}. {Besides, the blockchain type should also be determined according to the desired decentralisation level, for the subsequent identity verification process}~\citep{selected19}. {In the \textit{analysis} phase, finalising the governance decisions usually depends on the coordination of relevant entities via voting}~\citep{selected2, selected4, selected7}. {Afterwards, specific on-chain rules are arranged in the \textit{design} phase. For instance, Proof-of-Work and various incentive mechanisms are usually integrated to align the different interest of stakeholders and prevent uncooperative behaviors which may harm the platform}~\citep{selected3}. {Regarding the \textit{implementation} phase, forking is viewed as a special form to implement new functionalities in blockchain}~\citep{selected8, selected9, selected12}. {Further, in the \textit{testing} phase, implemented codes need to be deployed to testnet for a certain time period before officially in use}~\citep{selected23}.

{Subsequently, governance during the \textit{operation} of a blockchain usually happens when human interventions are required to deal with emergencies}~\citep{selected4}, {for instance, violations to laws}~\citep{selected25}, {software bugs}~\citep{selected34}, {or disputes between different entities}~\citep{selected32}. {Such interventions are regarded as ``exogenous governance", which provides formal mechanisms for stakeholders to express their ideas on the future direction of a blockchain platform}~\citep{selected7}. {In these cases, proposals are submitted to change the status quo and introduce new policies to blockchain, which then start a new epoch of development process from the \textit{planning} phase}~\citep{selected24, selected30, selected33, selected36}.

{Finally, five primary studies mention the \textit{termination} of a blockchain platform from the perspective of participating entities. If the requirements cannot be satisfied, an entity can always choose ``exit or voice". ``Voice" means proposing or voting for new improvement proposals, while ``exit" refers to leaving the blockchain}~\citep{selected3, selected4, selected8, selected15, selected22}.

\begin{figure*}[h!]
\begin{center}
\setlength{\fboxrule}{0.8pt}
\noindent

\fbox{%
\begin{minipage}{\textwidth}
\vspace{0.7em}
\footnotesize
\begin{center}
    \textbf{Insight from RQ4: When is blockchain governance applied?}\\
\end{center} 

\begin{quotation}

{\textbf{\textit{Termination phase}}: Discussion on the termination of a blockchain is missing. In particular, when a blockchain project fails, how to distribute the resources to protect the rights and interests of the involved parties is still under exploration.}

{\textbf{\textit{Blockchain ecosystem}}: The current studies mainly focus on the governance process of blockchain platform, while further refinement is needed to analyse the governance process of other governance objects in blockchain ecosystem (e.g., data, application, community).}

\end{quotation}
\vspace{0.7em}
\end{minipage}}
\end{center} 
\end{figure*}


\begin{table*}[!h]
\footnotesize
\centering
\caption{{RQ5: Who is involved in blockchain governance?}}
\label{who}
\begin{tabular}{p{0.1\columnwidth}p{0.2\columnwidth}p{0.6\columnwidth}}
\toprule

\multicolumn{2}{c}{\bf Stakeholders} & {\bf Primary studies} \\
\cmidrule(l){1-3}

\multicolumn{2}{l}{\multirow{2}{0.3\columnwidth}{\bf Project team}} & \citep{selected3, selected4, selected7, selected8, selected9, selected10, selected11, selected12, selected13, selected14, selected15, selected16, selected17, selected19, selected20, selected21, selected22, selected25, selected26, selected27, selected31, selected32, selected33, selected34, selected35, selected36, selected37}\\
\cmidrule(l){1-3}

& \multirow{2}{0.2\columnwidth}{Developer} & \citep{selected3, selected4, selected7, selected8, selected9, selected10, selected11, selected12, selected13, selected14, selected15, selected16, selected17, selected19, selected20, selected21, selected22, selected27, selected33, selected34, selected35, selected36, selected37} \\
\cmidrule(l){2-3}

& Foundation & \citep{selected3, selected7, selected14, selected15, selected19, selected22, selected26, selected31, selected35} \\
\cmidrule(l){2-3}

& Project lead & \citep{selected19, selected25, selected32} \\
\cmidrule(l){2-3}

& Legal professional & \citep{selected11} \\
\cmidrule(l){1-3}

\multicolumn{2}{l}{\multirow{2}{0.3\columnwidth}{\bf Node operator}} & \citep{selected1, selected2, selected3, selected4, selected6, selected7, selected8, selected9, selected12, selected13, selected15, selected16, selected17, selected19, selected20, selected21, selected22, selected23, selected25, selected27, selected29, selected30, selected32, selected33, selected34, selected35, selected36, selected37}\\
\cmidrule(l){1-3}

& \multirow{1}{0.2\columnwidth}{Block validator} & \citep{selected2, selected3, selected4, selected6, selected7, selected8, selected9, selected12, selected15, selected16, selected17, selected19, selected20, selected23, selected27, selected33, selected34, selected35, selected36, selected37} \\
\cmidrule(l){2-3}

& \multirow{1}{0.15\columnwidth}{Full node operator} & \multirow{1}{0.6\columnwidth}{\citep{selected1, selected3, selected7, selected8, selected12, selected13, selected15, selected16, selected20, selected21, selected25, selected29, selected32, selected33, selected34, selected37}} \\
\cmidrule(l){2-3}

& \multirow{1}{0.15\columnwidth}{Special node operator} & \multirow{2}{0.6\columnwidth}{\citep{selected2, selected9, selected22, selected30}} \\ \\
\cmidrule(l){1-3}

\multicolumn{2}{l}{\multirow{2}{0.3\columnwidth}{\bf User}} & \citep{selected2, selected3, selected4, selected7, selected8, selected9, selected12, selected13, selected14, selected15, selected16, selected17, selected19, selected21, selected22, selected23, selected24, selected30, selected31, selected34, selected35, selected37}\\
\cmidrule(l){1-3}

& Token holder & \citep{selected4, selected7, selected8, selected21, selected23, selected30, selected31, selected35, selected37} \\
\cmidrule(l){1-3}

\multicolumn{2}{l}{\bf Application provider} & \citep{selected3, selected4, selected7, selected8, selected11, selected12, selected14, selected15, selected16, selected19, selected27, selected35}\\
\cmidrule(l){1-3}

& Exchange & \citep{selected3, selected4, selected7, selected8, selected15, selected16, selected35} \\
\cmidrule(l){2-3}

& Custodian & \citep{selected3, selected4} \\
\cmidrule(l){2-3}

& Wallet provider & \multirow{1}{0.6\columnwidth}{\citep{selected3, selected4, selected7}} \\
\cmidrule(l){2-3}

& Bank & \citep{selected16} \\
\cmidrule(l){1-3}

\multicolumn{2}{l}{\bf Regulator} & \citep{selected1, selected4, selected5, selected6, selected7, selected10, selected16, selected17, selected18, selected19, selected25, selected26, selected27, selected28, selected32}\\
\cmidrule(l){1-3}

& Government & \citep{selected4, selected6, selected7, selected10, selected16, selected17, selected18, selected19, selected25, selected26, selected28, selected32} \\
\cmidrule(l){2-3}

& Auditor & \citep{selected1, selected25} \\
\cmidrule(l){2-3}

& Court & \citep{selected32} \\
\cmidrule(l){1-3}

\multicolumn{2}{l}{\bf Media} & \citep{selected3, selected4, selected19}\\
\cmidrule(l){1-3}

\multicolumn{2}{l}{\bf Researcher} & \citep{selected11}\\
\cmidrule(l){1-3}

\multicolumn{2}{l}{\bf Environmentalist} & \citep{selected7}\\

\bottomrule
\end{tabular}
\end{table*}

\subsection{RQ5: Who is involved in blockchain governance?}
The motivation of RQ5 is to identify the stakeholders who {are related to the governance of blockchain and how they are involved}. We present the identified stakeholders in Table \ref{who}.



\begin{itemize}
    \item \textit{Project team}: {{27} out of {37} primary studies consider project team as the key stakeholder in blockchain governance. In general, the project team is responsible for both technical implementations codified to the blockchain and other real-world arrangements}~\citep{selected15}. {First, developers maintain the code of a blockchain platform that any upgrades to the blockchain are implemented by them}~\citep{selected8}. {Specifically, three studies mention the concept of ``benevolent dictator", which indicates that the core developers (e.g. Nakamoto who invented Bitcoin) of a blockchain platform usually have more decision rights than the other roles to overcome emergency situations}~\citep{selected9, selected14, selected35}. {Secondly, foundations provide financial support to the invention and development process, hence, they can also affect the governance decisions of blockchain platforms}~\citep{selected7}. {According to Reijers et al.} \citep{selected22}, {the Ethereum Foundation had a significant influence in regards to the responses to the ``DAO attack".} In addition, three studies \citep{selected19, selected25, selected32} point out that the project team should select a person as the lead who is responsible and accountable for the team decisions and handles legal issues. Pelt et al. \citep{selected11} have a similar suggestion of hiring legal professionals for blockchain firms.
    

    
    \item \textit{Node operator}: {28} out of {37} primary studies view node operators as the main stakeholders in blockchain governance. {First, the mostly discussed node operators are block validators (also known as ``miners" in blockchain-based decentralised finance). They are responsible for the generation and inclusion of new blocks (i.e., the data entries of a blockchain), and ensure the security by participating in the consensus mechanism}~\citep{selected12, selected19}. {Secondly, full node operators maintain the storage of all historical ledger data, and they can decide to support the blockchain platform upgrades by installing the latest version}~\citep{selected7}. Moreover, special node operators with particular decision rights can be either predetermined in design or selected during operation \citep{selected2, selected9, selected22, selected30}. For example, Dash allows stakeholders who own more than 1000 Dash tokens to be the ``masternodes", who can vote for the future direction of Dash blockchain \citep{selected30}.

    
    \item \textit{User}: {In the governance of blockchain, users are critical as they eventually determine whether a blockchain platform can survive. For instance, when a hard fork occurs, a blockchian splits into two versions. Users need to decide which version they will continue to use}~\citep{selected8, selected24}. {Besides, users can provide feedback to the project team about encountered problems during the usage of the blockchain platform}~\citep{selected8}. {Furthermore, regarding the close connection between permissionless blockchain and decentralised finance, there are seven studies mentioning token holders in blockchain governance, as listed in Table} \ref{who}. {In certain blockchain platforms, token holders can submit proposals that outline potential directions while other token holders can vote for the proposal}~\citep{selected21, selected23, selected30}.

    
    
    \item \textit{Application provider}: {As blockchain-based applications are considered as a critical part of blockchain ecosystem, the application providers also participate in the governance of blockchain. For instance, financial systems may influence the underlying blockchain platforms of decentralised finance via interactions or competition}~\citep{selected7, selected16}. {Cryptocurrency exchanges also play a significant role that they need to decide which coin (token) to support in the case that the related blockchain platform conducts a hard fork}~\citep{selected8}. {Similar providers include custodian, wallet provider, and bank}~\citep{selected3, selected4, selected7, selected16}.

    
    \item \textit{Regulator}: {Regulators refer to governments, court, and third-party auditors, who 
    ensure that all decisions and activities within blockchain ecosystem comply with related laws and policies. Governments can affect the governance decisons made by other stakeholders by introducing new legal restrictions and constraints}~\citep{selected4, selected18}. {Examples include regulations of tax or against criminal behaviors like money laundering, where court is also involved}~\citep{selected16, selected26, selected32}. {Meanwhile, auditors are responsible to record all required information for future investigation}~\citep{selected25}.

    
    \item \textit{Indirect stakeholders}: {In addition to the above stakeholders, we categorise the following three types as ``indirect stakeholders", including media, researcher, and environmentalist. They can all influence or contribute to the blockchain governance decisions while do not participate in the operation of a blockchain platform. Media and environmentalist can both make social pressure which may affect stakeholders' decisions}~\citep{selected3, selected4, selected7, selected19}. Whilst, Pelt et al. \citep{selected11} mention that researchers contribute to the blockchain technology by conducting academic studies. Researchers usually do not count themselves as the stakeholders within the blockchain community, whilst their studies indeed help form governance models and frameworks for blockchain platforms.
    

\end{itemize}

\begin{figure*}[h!]
\begin{center}
\setlength{\fboxrule}{0.8pt}
\noindent

\fbox{%
\begin{minipage}{\textwidth}
\vspace{0.7em}
\footnotesize
\begin{center}
    \textbf{Insight from RQ5: Who is involved in blockchain governance?}\\
\end{center} 

\begin{quotation}

{\textbf{\textit{Mapping between stakeholders and governance dimensions:}} Although the primary studies provide a high-level view of the roles in blockchain governance, a mapping of each role with each of the governance dimensions (i.e., decision rights, accountability, and incentives) is still under studied. For instance, how different node operators are incentivised to contribute to the blockchain operation?}

{\textbf{\textit{Data subject:}} Blockchain provides an immutable data storage, whereby there may be conflicts between such immutability and data subjects' rights, which need to be considered in the governance of blockchain. Specifically, an individual should have the full control of his/her personally identifiable information.}

\end{quotation}
\vspace{0.7em}
\end{minipage}}
\end{center} 
\end{figure*}

\begin{table*}[!h]
\footnotesize
\centering
\caption{{RQ6: How is blockchain governance designed?}}
\label{how}
\begin{tabular}{p{0.02\columnwidth}p{0.05\columnwidth}p{0.05\columnwidth}p{0.25\columnwidth}p{0.548\columnwidth}}
\toprule

\multicolumn{4}{c}{\bf Governance mechanisms} & \multicolumn{1}{c}{\bf Primary studies} \\
\cmidrule(l){1-5}

\multirow{11}{0.02\columnwidth}{\rotatebox{90}{\bf Process mechanisms}} & \multicolumn{3}{l}{\multirow{1}{0.35\columnwidth}{\bf Improvement proposal}} & \multicolumn{1}{l}{\multirow{1}{0.548\columnwidth}{\citep{selected2, selected3, selected7, selected8, selected9, selected12, selected13, selected14, selected17, selected19, selected20, selected21, selected22, selected23, selected24, selected30, selected33, selected35, selected36, selected37}}} \\
\cmidrule(l){2-5}

& \multicolumn{3}{l}{\multirow{1}{0.35\columnwidth}{\bf Institutional oversight}} & \multicolumn{1}{l}{\multirow{1}{0.548\columnwidth}{\citep{selected1, selected4, selected6, selected8, selected13, selected16, selected17, selected18, selected25, selected26, selected31, selected32}}} \\ 
\cmidrule(l){2-5}

& & \multicolumn{2}{l}{\multirow{1}{0.25\columnwidth}{Law enforcement}} & \citep{selected4, selected8, selected16, selected18, selected25, selected26} \\
\cmidrule(l){3-5}

& & \multicolumn{2}{l}{\multirow{1}{0.25\columnwidth}{Role assignment}} & \citep{selected4, selected25, selected26} \\
\cmidrule(l){3-5}

& & \multicolumn{2}{l}{\multirow{1}{0.25\columnwidth}{Auditing}} & \citep{selected1, selected25} \\
\cmidrule(l){3-5}

& & \multicolumn{2}{l}{\multirow{1}{0.25\columnwidth}{Emergency stop}} & \citep{selected16, selected25} \\
\cmidrule(l){3-5}

& & \multicolumn{2}{l}{\multirow{1}{0.25\columnwidth}{Off-chain contract}} & \citep{selected16} \\
\cmidrule(l){2-5}


& \multicolumn{3}{l}{\multirow{1}{0.35\columnwidth}{\bf Testing environment}} & \citep{selected23, selected27, selected28} \\ 
\cmidrule(l){1-5}

\multirow{20}{0.02\columnwidth}{\rotatebox{90}{\bf Product mechanisms}} & \multicolumn{3}{l}{\multirow{1}{0.35\columnwidth}{\bf Voting}} & \multicolumn{1}{l}{\multirow{1}{0.548\columnwidth}{\citep{selected1, selected2, selected4, selected7, selected8, selected9, selected11, selected12, selected13, selected14, selected15, selected17, selected20, selected21, selected22, selected23, selected30, selected33, selected37}}} \\
\cmidrule(l){2-5}

& & \multicolumn{2}{l}{\multirow{1}{0.25\columnwidth}{Carbonvote}} & \citep{selected8} \\
\cmidrule(l){3-5}

& & \multicolumn{2}{l}{\multirow{1}{0.25\columnwidth}{Cross-chain voting}} & \citep{selected21} \\
\cmidrule(l){3-5}

& & \multicolumn{2}{l}{\multirow{1}{0.25\columnwidth}{Quadratic voting}} & \citep{selected37} \\
\cmidrule(l){3-5}

& & \multicolumn{2}{l}{\multirow{1}{0.25\columnwidth}{IP restriction}} & \citep{selected30} \\
\cmidrule(l){2-5}

& \multicolumn{3}{l}{\multirow{1}{0.35\columnwidth}{\bf Forking}} & \multicolumn{1}{l}{\multirow{1}{0.548\columnwidth}{\citep{selected3, selected4, selected8, selected9, selected11, selected12, selected13, selected14, selected15, selected20, selected22, selected24, selected25, selected34, selected35, selected36}}} \\
\cmidrule(l){2-5}

& \multicolumn{3}{l}{\multirow{1}{0.35\columnwidth}{\bf Consensus protocol}} & \multicolumn{1}{l}{\multirow{1}{0.548\columnwidth}{\citep{selected2, selected7, selected9, selected11, selected12, selected13, selected14, selected15, selected18, selected19, selected20, selected22, selected29, selected35}}} \\
\cmidrule(l){2-5}

& & \multicolumn{2}{l}{\multirow{1}{0.25\columnwidth}{Role election}} & \citep{selected2, selected9, selected19} \\
\cmidrule(l){2-5}

& \multicolumn{3}{l}{\multirow{1}{0.35\columnwidth}{\bf Incentive mechanism}} & \multicolumn{1}{l}{\multirow{1}{0.548\columnwidth}{\citep{selected2, selected3, selected7, selected9, selected11, selected12, selected14, selected15, selected19, selected20, selected23, selected30, selected31, selected34, selected37}}} \\
\cmidrule(l){2-5}

& & \multicolumn{2}{l}{\multirow{1}{0.25\columnwidth}{Token burner}} & \citep{selected23, selected30, selected37} \\
\cmidrule(l){2-5}

& \multicolumn{3}{l}{\multirow{1}{0.35\columnwidth}{\bf Platform modularisation}} & \multicolumn{1}{l}{\multirow{1}{0.548\columnwidth}{\citep{selected2, selected13, selected15, selected18, selected23}}} \\
\cmidrule(l){2-5}

& \multicolumn{3}{l}{\multirow{1}{0.35\columnwidth}{\bf Participation permission}} & \multicolumn{1}{l}{\multirow{1}{0.548\columnwidth}{\citep{selected11, selected14, selected19}}} \\
\cmidrule(l){2-5}

& \multicolumn{3}{l}{\multirow{1}{0.35\columnwidth}{\bf Transaction filter}} & \multicolumn{1}{l}{\multirow{1}{0.548\columnwidth}{\citep{selected6, selected12}}} \\

\bottomrule
\end{tabular}
\end{table*}

\subsection{RQ6: How is blockchain governance designed?}

The final research question is to explore actionable mechanisms for implementing blockchian governance. As illustrated in Table \ref{how}, the governance mechanisms can be classified into two categories: process mechanisms and product mechanisms. {Hereby, process mechanisms describe the steps of blockchain development via the governance meta-rules. Whilst, product mechanisms include the features of a blockchain, as the final outcomes of software development process.}


{The extracted process mechanisms are summarised as follows.}

\begin{itemize}
    \item \textit{Improvement proposal}: An improvement proposal is created to address unexpected exceptions and changing requirements. Two well-known examples are the Bitcoin improvement proposal and Ethereum improvement proposal \citep{selected2}. The project team can collect the feedback regarding the future roadmap of blockchain platform from other blockchain stakeholders (e.g., Ethereum Request for Comment \citep{selected13}) via offline seminars, online posts and mailing lists. Usually, there is a valid period for a proposal so that an expired proposal cannot be processed anymore \citep{selected2}.
    
    

    \item \textit{Institutional oversight}: 12 papers discuss institutional oversight for blockchain, which depends on the related governments and organisations. First, laws and policies for blockchain governance should be issued and enforced by the states. {Filippi et al.} \citep{selected4} {state that blockchain needs to adapt constitutional safeguards from centralised coordinating authorities, for instance, legally mandated mutability}~\citep{selected8, selected18},{ finance-related laws regarding the decentralised finance applications}~\citep{selected16, selected26}, {etc. Secondly, authorities can assign specific roles to participate in the governance of blockchain.} Atzori et al. \citep{selected6} propose a network where only the trust service providers, which are appointed by European governmental agencies, can become transaction validators. {Thirdly, in certain emergency cases, regulators should be capable to stop all on-chain business via extreme methods like blocking data transmission at network layer}~\citep{selected16, selected25}. In addition, auditing the blockchain historical data is necessary to ensure the legitimacy of on-chain business activities \citep{selected25, selected1}. Finally, as the smart contracts are deployed on-chain to achieve self-autonomy, where the codes are immutable, off-chain contracts or agreements are required in a business relationship, to flexibly make changes to the ``non-smart" parts \citep{selected16}.

    \item \textit{Testing environment}: Testnet and sandbox are two techniques providing separate running programs for testing new codes or functionalities without influencing the actual business activities in blockchain platforms \citep{selected23, selected27, selected28}.

\end{itemize}

{Extracted product mechanisms are listed as follows.}

\begin{itemize}
    \item \textit{Voting}: \textit{Voting} is commonly used {as a conflict resolution method to finalise governance decisions,} which can be held either on-chain  or off-chain. Specifically, node operators and users are the dominant roles in a voting process~\citep{selected2,selected8, selected9, selected12}, to support or veto particular on-chain activities (e.g., the acceptance of proposals~\citep{selected14, selected17, selected20}, audit of historical transactions\citep{selected1}, and the election of particular roles~\citep{selected7}). {Besides, different variants are proposed to ensure on-chain voting is viable and fair. Finck}~\citep{selected8} {discusses \textit{carbonvote} in which votes are counted regarding the tokens owned by voted individuals.} Fan et al. \citep{selected21} propose a \textit{cross-chain voting} scheme for new blockchain platforms via distributing tokens in other famous public blockchains, whereby token holders can vote for the new policies of this new blockchain. {Wright \citep{selected37} analyses quadratic voting, which can capture the preferences of votes via quadratically-increased tokens required for each additional vote.} {Mosley et al.} \citep{selected30} {mention that on-chain voting systems can ensure voting integrity via restricting IP spoofing. However, a major challenge of implementing voting in blockchain is that individuals lacking technical expertise can also make their votes}~\citep{selected13}. 
    
    \item \textit{Forking}: After the voting for an improvement proposal from all related stakeholders, {forking is conducted to implement accepted proposals. It consists of two types of upgrade: 1) soft forks refer to backward-compatible software upgrades to a blockchain platform; 2) hard forks mean backward-incompatible upgrades that all stakeholders need to discard the previous version}~\citep{selected3, selected8, selected9, selected12, selected22, selected25}. {Please note that there is also a special situation of forking: a subset of stakeholders are not satisfied with a particular governance decision, hence, they leave the current blockchain platform and then perform a hard fork to establish a separate blockchain instance as a means of veto}~\citep{selected4}.

    \item \textit{Consensus protocol}: {The consensus protocol specifies how participants behave when interacting with a blockchain platform. The on-chain protocol is autonomously executed by blockchain stakeholders}~\citep{selected13}. {Reijers et al.} \citep{selected22} {discuss that consensus protocols can be regarded as a shared concept of authority within a blockchain platform. Allen and Berg} \citep{selected7} {state that the consensus protocols in permissionless blockchains can maintain and protect the platform by coordinating self-interested behaviors.} Specifically, the definition of the authority, capability and responsibility of different roles is viewed as a significant part a in consensus protocol. For instance, in Dash blockchain, applying for ``Masternode" requires the ownership of at least 1000 dash tokens, to express the importance of a participant in the blockchain platform \citep{selected9}. Baudlet et al. \citep{selected2} adopt importance scores calculation in the election of ``Honorary Masternode". These special node operators can vote for improvement proposal. Kim \citep{selected29} proposes a consensus mechanism in which the honest nodes can collaborate to defend the attack from malicious nodes.
    
    \item \textit{Incentive mechanism}: {In addition to consensus protocol, the incentive mechanisms, along with inherent token distribution in permissionless blockchain platforms, is regarded as a motivational factor for stakeholders to participate in governance-related activities}~\citep{selected3, selected7, selected15, selected34}. {The embedded incentive mechanisms are usually based on game-theoretic insights, to encourage the operation of consensus protocol regarding the election of block validators}~\citep{selected2, selected9, selected12, selected20}. {Furthermore, incentive mechanisms also include preventing or punishing violations based on stakeholders' behaviors. For instance, the blockchain community can burn a service provider's staked tokens via voting, if the terms of service are not fulfilled}~\citep{selected23}. {In Dash blockchain, proposing an improvement amendment costs 5 Dash tokens to restrict the spams}~\citep{selected30}.

    \item \textit{Platform modularisation}: {Five studies mention that blockchain platforms need to perform modularisation to achieve flexible replacement of on-chain components. In particular, the movement from Proo-of-Work consensus protocol to Proof-of-Stake can adjust the allocation of decision rights, avoid energy consumption, while also addressing scalability issues}~\citep{selected2, selected15}. {The replacement of deployed smart contracts can improve the adaptability towards changes of application domains}~\citep{selected13}. {Similarly, the community can vote to adjust the infrastructure settings of blockchain platform (e.g., block size, block interval)}~\citep{selected23}. {Finally, government agencies should be allowed to edit blockchain platforms under certain conditions to ensure the conformance to laws}~\citep{selected18}.
    
    \item \textit{Participation permission}: In certain blockchain networks, especially permissioned blockchains, new users are verified before joining the systems \citep{selected11, selected14, selected19}.
    
    \item \textit{Transaction filter}: {Transactions are the data entries of a blockchain platform, while block validators are viewed having the capability to manually filter the transactions to ensure data source and quality}~\citep{selected6, selected12}.
\end{itemize}

Note that majority of the primary studies distinguish governance of permissioned blockchain from permissionless blockchain. The major difference between these two types of blockchain is that permissioned ones may have an authority dominating the blockchain platform \citep{selected11, selected26}. Consequently, governing a permissioned blockchain can refer to existing centralised decision-making models as only a few stakeholders are involved \citep{selected3}. {\textit{Participation permission} is usually applied in this type of blockchain platforms.} Meanwhile, the roles in a permissioned blockchain are more of assigned than elected. On the contrary, governance is more complex in permissionless blockchain where there is higher level of decentralisation \citep{selected8}. In this case, effective negotiation {and voting mechanisms} are significant for the stakeholders to reach consensus when making a decision.

{In addition to the above extracted mechanisms, seven studies propose governance frameworks for blockchain platforms, as checkbox to provide guidance for the future design and testing. Katina et al.} \citep{selected5} {propose seven interrelated elements (philosophy, theory, axiology, methodology, axiomatic, method and applications), while Allen and Berg} \citep{selected7} {provide a descriptive framework to understand exogenous and endogenous governance in blockchain. John and Pam} \citep{selected10} {and Pelt et al.} \citep{selected11} {both study on-chain and off-chain development processes to adopt governance. Beck et al.} \citep{selected14} {formulate a blockchain governance framework which is centered with the three dimensions of decision rights, accountability, and incentives in IT governance. Howell et al.} \citep{selected15} {focus on the membership and transacting relationships, and Werner et al.} \citep{selected31} {develop a taxonomy of platform governance for blockchain.}

\begin{figure*}[h!]
\begin{center}
\setlength{\fboxrule}{0.8pt}
\noindent

\fbox{%
\begin{minipage}{\textwidth}
\vspace{0.7em}
\footnotesize
\begin{center}
    \textbf{Insight from RQ6: How is blockchain governance designed?}\\
\end{center} 

\begin{quotation}

{\textit{\textbf{Decentralisation nature}}: Certain governance mechanisms are embedded in the design of blockchain regarding the inherent decentralisation nature. For instance, the design and implementation process of \textit{consensus protocol} and \textit{incentive mechanism} embody the product features of a blockchain.}

{\textit{\textbf{Sharding}}: Discussion of sharding technique is missing. Sharding determines how a blockchain platform is constructed, including how users are partitioned to different shards and how data is stored. Further, the governance of a whole blockchain ecosystem may be split into small pieces along with the shards.}

{\textit{\textbf{Source code management}}: The transparency degree of source code may affect the governance of blockchain. Currently, the code of many permissionless blockchain platforms (e.g., Bitcoin, Ethereum) are visible to the public, to accept contribution and also regulate the development process. Nevertheless, such openness allows the copy of code and hence, reduces the difficulty of \textit{forking} to compete with the original blockchain.}

\end{quotation}
\vspace{0.7em}
\end{minipage}}
\end{center} 
\end{figure*}

\section{Research Agenda}
\label{agenda}
{In this section, we provide a research agenda of this topic as future study directions based on the performed SLR and our insights.}

{\textbf{Propose High-level Governance Principles}: Future studies can propose governance principles to provide guidance to practitioners.} 

\begin{itemize}
    \item {How to ensure the blockchain ecosystem is compliant to legal regulations and ethical responsibilities? The governance of blockchain needs to comply with the legal regulations which set the minimum standards of human behaviors. Researchers need to analyse the regulations issued by different countries and how to codify the rules into blockchain platforms. Further, future studies can focus on how to preserve human values and even manage broader ethical responsibilities which denote the maximum standards of human behaviors. For instance, how to develop blockchain as a responsible technology via proper governance?}
    
    \item {What's the overall governance process throughout the blockchain ecosystem? The lifecycles can be extended to ecosystem-level to include other blockchain governance objects (e.g., data, platform, application, community).}
    
    \item {What decision rights are allocated to different stakeholders? How to ensure they are accountable? How are they incentivised? The mapping between blockchain stakeholders and their decision rights, accountability, and incentives is significant to derive the responsibility assignment matrix from the governance perspective.}
\end{itemize}


{\textbf{Process Best Governance Practices}: Further research can investigate the actionable governance solutions for blockchain development.}

\begin{itemize}
    \item {What are the reusable process and patterns for blockchain governance? For instance, on-chain autonomous governance is dependent on the codified consensus protocol and incentive mechanism. On one hand, deployed consensus protocol should align with the decentralisation level of a blockchain platform. On the other hand, the rewards and sanctions involved in the incentive mechanism embodies how the project team value the contribution of stakeholders. Future studies can be carried out in these two directions to facilitate effective on-chain governance.}
    
    \item {How can blockchain systems be designed to enable governance? A governance-driven software architecture for blockchain-based systems can help practitioner operationalise governance methods for the future design and development of blockchain-based systems.}
    
    \item {How to evaluate the governance of blockchain? A maturity model can serve as an internal benchmarking tool to understand the degree of governance in a certain blockchain.}  

\end{itemize}

\section{Threats To Validity}
\label{threats}

{According to SLR guidelines} \citep{keele2007guidelines, threats_to_validity}, {threats to validity may be introduced during the review process and affect the whole study. In this section, we identify the threats to four different types of validity (i.e., construct validity, internal validity, external validity, and conclusion validity), and discuss the strategies we adopted to minimise their influences.}

{\textbf{Construct Validity}: The threat to construct validity is the incompleteness of search strings and its possible effect of insufficiency on study collection. Originally, we intended to include the synonyms of ``governance" in the search strings, such as ``rule", ``administration", ``supervision", ``management", etc. These words indeed enriched the search results, nevertheless, we found that these synonyms brought ambiguity into the topic of blockchain governance. Consequently, to avoid misunderstanding, we chose a conservative strategy that only using two cognates of ``governance" (i.e., govern, governing) as keywords.}

{\textbf{Internal Validity}: Publication bias and inadequate size of samples may introduce threats to internal validity. Publication bias means that most of studies with evaluation have positive results over negative results. Papers that have significant or positive results may have a higher possibility to be accepted than those with null or negative results, which might lead to a biased publication tendency. We carefully reviewed all included studies to mitigate the effects of this threat. Regarding the inadequate size of primary studies, we included book chapters and papers from online archives as they are all searched by the source of Google Scholar. Four researchers reviewed these studies to ensure that they adhered to the predefined protocol.}

{\textbf{External Validity}: Threat to external validity refers to the restricted time span in our protocol. We searched papers published 1 Jan 2008 to 3 Nov 2020. We chose 2008 as Bitcoin was proposed in this year, which has boosted the development of blockchain. However, the history of blockchain technology can trace back to 1979}~\citep{blockchain_origin}. {The excluded studies published before 2008 may affect the generalisability of our result.}

{\textbf{Conclusion Validity}: Threats may be introduced due to the bias in study selection and data extraction phases. During the inclusion and exclusion of searched studies, we found that a set of studies focusing on cryptocurrency governance, which is understandable as decentralised finance is the first well-known application of blockchain technology. We excluded some of them based on our predefined protocol, if the study is mostly discussing financial issues without proper design or process solution, tools or toolkit for the governance of blockchain. Bias in data extraction phase happens as two researchers conducted the extraction from assigned studies and cross-validated each other's results. However, the extraction situation may vary, considering the different experiences and knowledge reserves of the researchers. For instance, whether framework development should be regarded as a governance solution for blockchain and reported in the paper. Another two independent researchers were consulted to reduce the biased influences in this procedure.}

\section{Conclusion}
\label{conclusion}

The importance of proper governance for blockchain is increasingly recognised in both industry and academia. To understand the state-of-art in this domain, a systematic literature review is performed in this paper. We collected 1061 papers and selected {37} of them as primary studies for this in-depth review based on a predefined protocol. After data extraction and synthesis, we present our study results and insights in this paper, aiming to provide guidance to researchers and practitioners for future studies in blockchain governance. 

The primary studies reveal that the current main goals of governance are the adaptability and upgradability of blockchain and governance methods mainly focus on the planning and operation of blockchain platform. From the extracted data, a common combination of governance mechanisms is the series of \textit{improvement proposal}, \textit{voting}, and \textit{forking} to upgrade the blockchain platform. {Meanwhile, the design and operation of \textit{consensus protocol} and \textit{incentive mechanism} are significant to on-chain autonomous governance by regulating stakeholders' behaviors. As blockchian governance is an ongoing topic, we provide the research agenda to future directions, including refining the lifecycle of different governance objects, studying the mapping between stakeholders and their decision rights, accountability, and incentives, while also analysing the legal regulations and ethical responsibilities. We plan to propose a governance framework and a design pattern catalogue for blockchain governance in the future.}

\section{Appendix}
\label{appendix}

{Please see Google doc\footnote{\url{https://docs.google.com/document/d/1ZAMWbTj1OIqkIOtafqli6sa73oqhR1u5/edit?usp=sharing&ouid=101956301517606860290&rtpof=true&sd=true}} for SLR protocol}, Table~\ref{primarystudies} and Google sheet\footnote{\url{https://docs.google.com/spreadsheets/d/1X4mqQLibO1bhsCcDebpGY0QCSjLai-Yl/edit?usp=sharing&ouid=101956301517606860290&rtpof=true&sd=true}} for extracted data. 

{
\footnotesize
\centering
\setlength{\tabcolsep}{2pt}
\begin{spacing}{1}
\begin{longtable}{p{0.05\columnwidth}p{0.06\columnwidth}p{0.52\columnwidth}p{0.11\columnwidth}p{0.06\columnwidth}p{0.1\columnwidth}}
\caption{Selected primary studies.}
\label{primarystudies}\\

\toprule
\bf{ID} &
\bf{Ref.} &
{\bf Title} &
\bf{Author} &
\bf{Year} &
\bf{Source} 
\\
\midrule
\endfirsthead

\multicolumn{4}{c}{{\bfseries \tablename\ \thetable{} -- continued from previous page}} \\
\toprule
\bf{ID} &
\bf{Ref.} &
{\bf Title} &
\bf{Author} &
\bf{Year} &
\bf{Source} 
\\
\midrule
\endhead

\midrule
\multicolumn{4}{r}{{Continued on next page}} \\ 
\bottomrule
\endfoot

\bottomrule
\endlastfoot

\multirow{1}{0.05\columnwidth}{S1} & \multirow{1}{0.06\columnwidth}{\citep{selected1}} & An Auditable and Secure Model for Permissioned Blockchain & \multirow{1}{0.11\columnwidth}{Bao et al.} & \multirow{1}{0.06\columnwidth}{2019} & \multirow{1}{0.1\columnwidth}{ACM} \\
\cmidrule{1-6}

\multirow{2}{0.05\columnwidth}{S2} & \multirow{2}{0.06\columnwidth}{\citep{selected2}} & The Best of Both Worlds: A New Composite Framework Leveraging PoS and PoW for Blockchain Security and Governance & \multirow{2}{0.11\columnwidth}{Baudlet et al.} & \multirow{2}{0.06\columnwidth}{2020} & \multirow{2}{0.1\columnwidth}{IEEE} \\
\cmidrule{1-6}

\multirow{2}{0.05\columnwidth}{S3} & \multirow{2}{0.06\columnwidth}{\citep{selected3}} & Bitcoin Governance as a Decentralized Financial Market Infrastructure & \multirow{2}{0.11\columnwidth}{Nabilou} & \multirow{2}{0.06\columnwidth}{2020} & \multirow{2}{0.1\columnwidth}{Google Scholar} \\
\cmidrule{1-6}

\multirow{2}{0.05\columnwidth}{S4} &
\multirow{2}{0.06\columnwidth}{\citep{selected4}} &
Blockchain as a Confidence Machine: The Problem of Trust \& Challenges of Governance &
\multirow{2}{0.11\columnwidth}{Filippi et al.} &
\multirow{2}{0.06\columnwidth}{2020} &
\multirow{2}{0.1\columnwidth}{Science Direct} \\
\cmidrule{1-6}

\multirow{2}{0.05\columnwidth}{S5} &
\multirow{2}{0.06\columnwidth}{\citep{selected5}} &
\multirow{2}{0.4\columnwidth}{Blockchain Governance} &
\multirow{2}{0.11\columnwidth}{Katina et al.} &
\multirow{2}{0.06\columnwidth}{2019} &
\multirow{2}{0.1\columnwidth}{Google Scholar} \\ \\
\cmidrule{1-6}

\multirow{2}{0.05\columnwidth}{S6} &
\multirow{2}{0.06\columnwidth}{\citep{selected6}} &
Blockchain Governance and The Role of Trust Service Providers: The TrustedChain Network &
\multirow{2}{0.11\columnwidth}{Atzori} &
\multirow{2}{0.06\columnwidth}{2018} &
\multirow{2}{0.1\columnwidth}{Google Scholar} \\
\cmidrule{1-6}

\multirow{2}{0.05\columnwidth}{S7} &
\multirow{2}{0.06\columnwidth}{\citep{selected7}} &
Blockchain Governance: What We can Learn from the Economics of Corporate Governance &
\multirow{2}{0.11\columnwidth}{Allen and Berg} &
\multirow{2}{0.06\columnwidth}{2020} &
\multirow{2}{0.1\columnwidth}{Google Scholar} \\
\cmidrule{1-6}

\multirow{2}{0.05\columnwidth}{S8} &
\multirow{2}{0.06\columnwidth}{\citep{selected8}} &
Blockchain Governance (from \textit{Blockchain Regulation and Governance in Europe}) &
\multirow{2}{0.11\columnwidth}{Finck} &
\multirow{2}{0.06\columnwidth}{2018} &
\multirow{2}{0.1\columnwidth}{Google Scholar} \\
\cmidrule{1-6}

\multirow{2}{0.05\columnwidth}{S9} &
\multirow{2}{0.06\columnwidth}{\citep{selected9}} &
Comparison and Analysis of Governance Mechanisms Employed by Blockchain-Based Distributed Autonomous Organizations &
\multirow{2}{0.11\columnwidth}{DiRose and Mansouri} &
\multirow{2}{0.06\columnwidth}{2018} &
\multirow{2}{0.1\columnwidth}{IEEE} \\
\cmidrule{1-6}

\multirow{2}{0.05\columnwidth}{S10} &
\multirow{2}{0.06\columnwidth}{\citep{selected10}} &
\multirow{2}{0.52\columnwidth}{Complex Adaptive Blockchain Governance} &
\multirow{2}{0.11\columnwidth}{John and Pam} &
\multirow{2}{0.06\columnwidth}{2018} &
\multirow{2}{0.1\columnwidth}{Google Scholar} \\ \\
\cmidrule{1-6}

\multirow{2}{0.05\columnwidth}{S11} &
\multirow{2}{0.06\columnwidth}{\citep{selected35}} &
\multirow{2}{0.4\columnwidth}{The DAO: a million dollar lesson in blockchain governance} &
\multirow{2}{0.11\columnwidth}{Santos and Kostakis} &
\multirow{2}{0.06\columnwidth}{2018} &
\multirow{2}{0.1\columnwidth}{Google Scholar} \\ \\
\cmidrule{1-6}

\multirow{2}{0.05\columnwidth}{S12} &
\multirow{2}{0.06\columnwidth}{\citep{selected11}} &
Defining Blockchain Governance: A Framework for Analysis and Comparison &
\multirow{2}{0.11\columnwidth}{Pelt et al.} &
\multirow{2}{0.06\columnwidth}{2020} &
\multirow{2}{0.1\columnwidth}{Google Scholar} \\
\cmidrule{1-6}

\multirow{2}{0.05\columnwidth}{S13} &
\multirow{2}{0.06\columnwidth}{\citep{selected12}} &
Distributed Governance in Multi-sided Platforms: A Conceptual Framework from Case: Bitcoin &
\multirow{2}{0.11\columnwidth}{Mattila and Sepp{\"a}l{\"a}} &
\multirow{2}{0.06\columnwidth}{2018} &
\multirow{2}{0.1\columnwidth}{Springer Link} \\
\cmidrule{1-6}

\multirow{2}{0.05\columnwidth}{S14} &
\multirow{2}{0.06\columnwidth}{\citep{selected36}} &
Egalitarian Society or Benevolent Dictatorship: The State of Cryptocurrency Governance &
\multirow{2}{0.11\columnwidth}{Azouvi et al.} &
\multirow{2}{0.06\columnwidth}{2018} &
\multirow{2}{0.1\columnwidth}{Springer Link} 
\\

\multirow{2}{0.05\columnwidth}{S15} &
\multirow{2}{0.06\columnwidth}{\citep{selected13}} &
Governance Challenges of Blockchain and Decentralized Autonomous Organizations &
\multirow{2}{0.11\columnwidth}{Rikken et al.} &
\multirow{2}{0.06\columnwidth}{2019} &
\multirow{2}{0.1\columnwidth}{Google Scholar} \\
\cmidrule{1-6}

\multirow{2}{0.05\columnwidth}{S16} &
\multirow{2}{0.06\columnwidth}{\citep{selected14}} &
Governance in the Blockchain Economy: A Framework and Research Agenda &
\multirow{2}{0.11\columnwidth}{Beck et al.} &
\multirow{2}{0.06\columnwidth}{2018} &
\multirow{2}{0.1\columnwidth}{Google Scholar} \\
\cmidrule{1-6}

\multirow{2}{0.05\columnwidth}{S17} &
\multirow{2}{0.06\columnwidth}{\citep{selected15}} &
Governance of Blockchain and Distributed Ledger Technology Projects &
\multirow{2}{0.11\columnwidth}{Howell et al.} &
\multirow{2}{0.06\columnwidth}{2019} &
\multirow{2}{0.1\columnwidth}{Google Scholar} \\
\cmidrule{1-6}

\multirow{2}{0.05\columnwidth}{S18} &
\multirow{2}{0.06\columnwidth}{\citep{selected16}} &
\multirow{2}{0.4\columnwidth}{The Governance of Blockchain Financial Networks} &
\multirow{2}{0.11\columnwidth}{Paech} &
\multirow{2}{0.06\columnwidth}{2017} &
\multirow{2}{0.1\columnwidth}{Google Scholar} \\ \\
\cmidrule{1-6}

\multirow{2}{0.05\columnwidth}{S19} &
\multirow{2}{0.06\columnwidth}{\citep{selected17}} &
The Governance of Blockchain Systems from an Institutional Perspective, a Matter of Trust or Control &
\multirow{2}{0.11\columnwidth}{Meijer and Ubacht} &
\multirow{2}{0.06\columnwidth}{2018} &
\multirow{2}{0.1\columnwidth}{ACM} \\
\cmidrule{1-6}

\multirow{2}{0.05\columnwidth}{S20} &
\multirow{2}{0.06\columnwidth}{\citep{selected18}} &
Governing the Use of Blockchain and Distributed Ledger Technologies: Not One-Size-Fits-All &
\multirow{2}{0.11\columnwidth}{Trump et al.} &
\multirow{2}{0.06\columnwidth}{2018} &
\multirow{2}{0.1\columnwidth}{IEEE} \\
\cmidrule{1-6}

\multirow{2}{0.05\columnwidth}{S21} &
\multirow{2}{0.06\columnwidth}{\citep{selected19}} &
The Internal and External Governance of Blockchain-based Organizations: Evidence from Cryptocurrencies (from \textit{Bitcoin and Beyond}) &
\multirow{2}{0.11\columnwidth}{Hsieh et al.} &
\multirow{2}{0.06\columnwidth}{2017} &
\multirow{2}{0.1\columnwidth}{Google Scholar} \\
\cmidrule{1-6}

\multirow{2}{0.05\columnwidth}{S22} &
\multirow{2}{0.06\columnwidth}{\citep{selected20}} &
\multirow{2}{0.52\columnwidth}{The Invisible Politics of Bitcoin: Governance Crisis of a Decentralised Infrastructure} &
\multirow{2}{0.11\columnwidth}{Filippi and Loveluck} &
\multirow{2}{0.06\columnwidth}{2016} &
\multirow{2}{0.1\columnwidth}{Google Scholar} \\ \\
\cmidrule{1-6}

\multirow{2}{0.05\columnwidth}{S23} &
\multirow{2}{0.06\columnwidth}{\citep{selected21}} &
MULTAV: A Multi-chain Token Backed Voting Framework for Decentralized Blockchain Governance &
\multirow{2}{0.11\columnwidth}{Fan et al.} &
\multirow{2}{0.06\columnwidth}{2020} &
\multirow{2}{0.1\columnwidth}{Springer Link} \\
\cmidrule{1-6}

\multirow{2}{0.05\columnwidth}{S24} &
\multirow{2}{0.06\columnwidth}{\citep{selected22}} &
Now the Code Runs Itself: On-Chain and Off-Chain Governance of Blockchain Technologies &
\multirow{2}{0.11\columnwidth}{Reijers et al.} &
\multirow{2}{0.06\columnwidth}{2018} &
\multirow{2}{0.1\columnwidth}{Springer Link} \\
\cmidrule{1-6}

\multirow{2}{0.05\columnwidth}{S25} &
\multirow{2}{0.06\columnwidth}{\citep{selected23}} &
Ping-Pong Governance: Token Locking for Enabling Blockchain Self-governance &
\multirow{2}{0.11\columnwidth}{Merrill et al.} &
\multirow{2}{0.06\columnwidth}{2020} &
\multirow{2}{0.1\columnwidth}{Springer Link} \\
\cmidrule{1-6}

\multirow{2}{0.05\columnwidth}{S26} &
\multirow{2}{0.06\columnwidth}{\citep{selected24}} &
\multirow{2}{0.52\columnwidth}{The Political Economy of Blockchain Governance} &
\multirow{2}{0.11\columnwidth}{Lee et al.} &
\multirow{2}{0.06\columnwidth}{2020} &
\multirow{2}{0.1\columnwidth}{Google Scholar} \\ \\
\cmidrule{1-6}

\multirow{2}{0.05\columnwidth}{S27} &
\multirow{2}{0.06\columnwidth}{\citep{selected37}} &
\multirow{2}{0.52\columnwidth}{Quadratic Voting and Blockchain Governance} &
\multirow{2}{0.11\columnwidth}{Wright} &
\multirow{2}{0.06\columnwidth}{2019} &
\multirow{2}{0.1\columnwidth}{Google Scholar} \\ \\
\cmidrule{1-6}

\multirow{2}{0.05\columnwidth}{S28} &
\multirow{2}{0.06\columnwidth}{\citep{selected25}} &
\multirow{2}{0.52\columnwidth}{Regulating Blockchain, DLT and Smart Contracts: a Technology Regulator’s Perspective} &
\multirow{2}{0.11\columnwidth}{Joshua et al.} &
\multirow{2}{0.06\columnwidth}{2020} &
\multirow{2}{0.1\columnwidth}{Springer Link} \\ \\
\cmidrule{1-6}

\multirow{2}{0.05\columnwidth}{S29} &
\multirow{2}{0.06\columnwidth}{\citep{selected26}} &
\multirow{2}{0.52\columnwidth}{Regulatory Issues in Blockchain Technology} &
\multirow{2}{0.11\columnwidth}{Yeoh} &
\multirow{2}{0.06\columnwidth}{2017} &
\multirow{2}{0.1\columnwidth}{Google Scholar} \\ \\
\cmidrule{1-6}

\multirow{2}{0.05\columnwidth}{S30} &
\multirow{2}{0.06\columnwidth}{\citep{selected27}} &
Sandbox for Minimal Viable Governance of Blockchain Services and DAOs: CLAUDIA &
\multirow{2}{0.11\columnwidth}{Arribas et al.} &
\multirow{2}{0.06\columnwidth}{2020} &
\multirow{2}{0.1\columnwidth}{Springer Link} \\
\cmidrule{1-6}

\multirow{2}{0.05\columnwidth}{S31} &
\multirow{2}{0.06\columnwidth}{\citep{selected28}} &
Sandboxes and Testnets as ``Trading Zones" for Blockchain Governance &
\multirow{2}{0.11\columnwidth}{Kera} &
\multirow{2}{0.06\columnwidth}{2020} &
\multirow{2}{0.1\columnwidth}{Springer Link} \\
\cmidrule{1-6}

\multirow{2}{0.05\columnwidth}{S32} &
\multirow{2}{0.06\columnwidth}{\citep{selected29}} &
\multirow{2}{0.52\columnwidth}{Strategic Alliance for Blockchain Governance Game} &
\multirow{2}{0.11\columnwidth}{Kim} &
\multirow{2}{0.06\columnwidth}{2020} &
\multirow{2}{0.1\columnwidth}{Google Scholar} \\ \\
\cmidrule{1-6}

\multirow{2}{0.05\columnwidth}{S33} &
\multirow{2}{0.06\columnwidth}{\citep{selected30}} &
Towards a Systematic Understanding of Blockchain Governance in Proposal Voting: A Dash Case Study &
\multirow{2}{0.11\columnwidth}{Mosley et al.} &
\multirow{2}{0.06\columnwidth}{2020} &
\multirow{2}{0.1\columnwidth}{Google Scholar} \\
\cmidrule{1-6}

\multirow{2}{0.05\columnwidth}{S34} &
\multirow{2}{0.06\columnwidth}{\citep{selected31}} &
Towards a Taxonomy for Governance Mechanisms of Blockchain-based Platforms &
\multirow{2}{0.11\columnwidth}{Werner et al.} &
\multirow{2}{0.06\columnwidth}{2020} &
\multirow{2}{0.1\columnwidth}{Google Scholar} \\
\cmidrule{1-6}

\multirow{2}{0.05\columnwidth}{S35} &
\multirow{2}{0.06\columnwidth}{\citep{selected32}} &
\multirow{2}{0.52\columnwidth}{Towards Governance and Dispute Resolution for DLT and Smart Contracts} &
\multirow{2}{0.11\columnwidth}{Erbguth and Morin} &
\multirow{2}{0.06\columnwidth}{2018} &
\multirow{2}{0.1\columnwidth}{IEEE} \\ \\
\cmidrule{1-6}

\multirow{2}{0.05\columnwidth}{S36} &
\multirow{2}{0.06\columnwidth}{\citep{selected33}} &
Why Blockchains need the Law Secondary Rules as the Missing Piece of Blockchain Governance &
\multirow{2}{0.11\columnwidth}{Crepaldi} &
\multirow{2}{0.06\columnwidth}{2019} &
\multirow{2}{0.1\columnwidth}{ACM} \\
\cmidrule{1-6}

\multirow{2}{0.05\columnwidth}{S37} &
\multirow{2}{0.06\columnwidth}{\citep{selected34}} &
\multirow{2}{0.52\columnwidth}{Why do Public Blockchains Need Formal and Effective Internal Governance Mechanisms} &
\multirow{2}{0.11\columnwidth}{Yeung and Galindo} &
\multirow{2}{0.06\columnwidth}{2019} &
\multirow{2}{0.1\columnwidth}{Google Scholar} \\ \\

\end{longtable}

\end{spacing}
}




\end{document}